\documentclass[fleqn,usenatbib,onecolumn]{mnras}

\usepackage[T1]{fontenc}
\usepackage{ae,aecompl}
\usepackage{bm}
\usepackage{graphicx}    

\usepackage{amsmath}	
\usepackage{amssymb}	
\usepackage{txfonts}

\DeclareMathOperator{\sgn}{sgn}

\title[Orbital dynamics of two circumbinary planets]{Orbital dynamics of two circumbinary planets around misaligned eccentric binaries}

\author[Chen et al.]{Cheng Chen$^1$\thanks{Email: chenc21@unlv.nevada.edu}, Stephen H. Lubow$^2,$ and Rebecca G. Martin$^1$
\\ $^1$Department of Physics and Astronomy,  University of Nevada, Las Vegas, 4505 South Maryland Parkway, Las Vegas, NV 89154, USA 
\\ $^{2}$Space Telescope Science Institute, 3700 San Martin Drive, Baltimore, MD 21218, USA\\}

\date{Accepted XXX. Received YYY; in original form ZZZ}

\pubyear{2020}

\begin{document}
\label{firstpage}
\pagerange{\pageref{firstpage}--\pageref{lastpage}} 
\maketitle

\begin{abstract}
We investigate the orbital dynamics of circumbinary planetary systems with two planets around a circular or eccentric orbit binary. The orbits of the two planet are initially circular and coplanar to each other, but misaligned with respect to the binary orbital plane. The binary-planet and planet-planet interactions result in complex planet tilt oscillations. We use analytic models and numerical simulations to explore the effects of various values of the planet semi-major axes, binary eccentricity, and initial inclination. Around a circular orbit binary, secular tilt oscillations are driven by planet-planet interactions and are periodic. In  that case, planets undergo mutual libration  if close together and circulation if far apart with an abrupt transition at a critical separation. Around an eccentric orbit binary, secular tilt oscillations are driven by both planet-planet interactions and binary-planet interactions. These oscillations generally display more than one frequency and are generally not periodic. The transition from mutual planet libration to circulation is not sharp and there is a range of separations for which the planets are on  orbits that are sometimes mutually librating and sometimes circulating. In addition, at certain separations, there are resonances for which tilt oscillations are complicated but periodic. 
For planets that are highly misaligned with respect to an eccentric orbit binary, there are stationary (non-oscillating) tilt configurations that are generalisations of polar configurations for the single planet case. Tilt oscillations of highly inclined planets occur for initial tilts that depart from the stationary configuration.
\end{abstract}


\section{Introduction}


Circumbinary discs are the birthplaces for circumbinary planets and so planets likely form with the same initial orbital properties as the discs. While planet formation very close to the binary may be suppressed, formation farther out may proceed in a similar way to around a single star \citep[e.g.][]{Moriwaki2004, Marzari2008}. Recent observations show that misaligned circumbinary discs are common around young binary systems  \citep[e.g.,][]{Chiang2004,Winn2004,Capelo2012,Kennedy2012,Brinch2016,Aly2018,Kennedy2019}. Although about 68\% of short period binaries (period $<20\,\rm days $) have discs that are coplanar with the binary orbital plane (with disc inclination relative to the binary of $\leq$ 3$^{\circ}$), longer orbital period binaries display a wider range of disc inclinations and binary eccentricities \citep{Czekala2019}.

The formation of misaligned discs may be due to chaotic accretion  \citep{clarke1993,Bate2003,Bate2018} or stellar flybys \citep{Cuello2019b,Nealon2020, Ma2020}. Protoplanetary discs typically nodally precess as a solid body if the disc is sufficiently narrow and and warm \citep{Larwoodetal1996,LP1997}. During the nodal preccesion, the angular momentum vector of the disc precesses about the angular momentum vector of the binary \citep[e.g.,][]{PT1995}. Dissipation due to the viscosity in the disc leads to tilt evolution \citep[e.g.,][]{Nixonetal2011a,Martin2017} and the disc moves towards either coplanar alignment with the orbital plane of the binary \citep{PT1995, PL1995, Lubow2000, Nixonetal2011a, Nixon2012, Kingetal2013, Facchinietal2013, Lodato2013, Foucart2013} or to a polar configuration where the disc angular momentum aligns to the binary eccentricity vector, perpendicular to the binary angular momentum vector \citep{Aly2015,Martin2017,MartinandLubow2018b,Lubow2018,Zanazzi2018,Smallwood2020}. 
In the polar case, planets may form in a highly misaligned configuration with respect
to the binary.
In addition, for a sufficiently extended disc, the disc lifetime may be longer than the alignment timescale.
Consequently, circumbinary planets with non-zero inclinations may form in misaligned discs or even polar discs.

Although misaligned circumbinary planets are expected to form in binaries with longer orbital periods, they are harder to detect by the transit method for two reasons. First, the probability of a transit is smaller than for coplanar systems and, second, because the orbital period of the planet is longer. To date, only 12 circumbinary planets have been detected by Kepler \citep{Doyle2011, Welsh2012, Orosz2012a, Orosz2012b, Kostov2014, Welsh2015, Li2016, Kostov2016, Socia2020} and they are all close to coplanar with the binary orbital plane. The coplanarity is likely a selection effect because the Kepler binaries have short orbital periods \citep{Czekala2019,MartinandLubow2019}. Eclipse timing variations of the binary is a better method to to distinguish inclined planets from coplanar planets \citep{Zhang2019}. In addition, there are several ways to detect circumbinary planets such as transit timing variations and transit duration variation \citep{Windemuth2019}, microlensing \citep{Bennett2016,Luhn2016}, and astrometry \citep{Sahlmann2015}.

Kepler-47 is currently the only binary system that has been observed to have multiple  circumbinary planets. The binary orbit is nearly circular ($e < 0.03$) and the three Neptune-size planets are close to coplanar to the orbital plane of the binary \citep{Orosz2012a, Orosz2012b, Kostov2013}. All of the circumbinary planets detected by Kepler are around  binaries with low eccentricity, except for Kepler-34b which has a host binary  eccentricity of 0.52 \citep{Welsh2012, Kley2015}. Low-mass, short-period binaries have stronger stellar tidal dissipation of their eccentricities as the two stars approach tidal locking \citep{Raghavan2010}. On the other hand, recent observations by {\it Transiting Exoplanet Survey  Satellite} (TESS)  have revealed the binary 1SWASPJ011351.29+314909.7 which has binary eccentricity 0.3098 \citep{Swayne2020} and a coplanar circumbinary planet, TOI-1338~b  which orbits a binary with an eccentricity of 0.156 and mass ratio $M_2/M_1= 0.217$ \citep{Hodzic2020,Kostov2020}. We expect that misaligned circumbinary planets around eccentric orbit binaries will be found in the future.

The angular momentum vector of a misaligned (massless) test particle orbiting around a circular orbit binary precesses around the binary angular momentum vector \citep{Farago2010}. 
These are  circulating orbits with respect to the binary in which the longitude of the ascending node fully circulates over 360$^\circ$ during the nodal precession. The behaviour is more complex for a binary with a non-zero eccentricity. In this case the angular momentum vector of a test particle with a sufficiently large initial orbital inclination 
 can precess about the binary eccentricity vector.
During this process, the particle orbit undergoes tilt oscillations and libration in which the longitude of the ascending node oscillates in a limited  range  \citep{Verrier2009,Farago2010,Doolin2011,Naoz2017,Vinson2018, deelia2019}. These are polar librating orbits. The minimum inclination (critical angle) required for polar librating orbits decreases as the binary eccentricity increases. Therefore, around a highly eccentric binary,  even an initially small inclination test particle orbit can librate.
 
The dynamics of a circumbinary planet around an eccentric orbit binary are also affected by the mass of the planet \citep{Chen20192}. For a misaligned non-zero mass planet orbiting around an eccentric binary, the critical inclination for the planet to librate depends on the binary eccentricity and the angular momentum ratio of the planet to the binary. The angle of the stationary inclination, or polar alignment, occurs at less than 90$^{\circ}$ if the planet is massive \citep{Farago2010, Lubow2018, Zanazzi2018, Chen20192, MartinandLubow2019}. Furthermore, in \citet{Chen20192} we found that the angular momentum transfer between the binary and a massive planet can be significant and the binary eccentricity can oscillate during the nodal libration. The  numerical simulation results are consistent with the analytic model in \citet{MartinandLubow2019}.

Multiple planets around a binary system can interact with each other, as well as the binary. Previous studies have investigated these interactions for planetary systems that are coplanar to the binary orbit. For example,  a secular apsidal resonance between the binary and the outer planet may be triggered due to the inner planet accelerating the apsidal precession rate of the binary \citep{Andrade-Ines2018}. The coplanar planet-planet resonances interact with the binary and result in overlapping mean motion resonances. Consequently, the orbits of circumbinary planets may be unstable. Planets can then be ejected or collide with the binary due to interactions with mean motion resonances \citep{Sutherland2016,Sutherland2019}. In this work, we focus on the orbits of planets that are non-coplanar to the binary orbit but initially coplanar to each other. For a misaligned system, an interacting inclined planet and disc around one component of a binary undergo secular tilt oscillations  \citep{Lubow2016}.  Two circumbinary planets may also have a similar interaction. To explore secular tilt oscillations interactions of circumbinary systems of two inclined planets, we develop analytic and numerical model.


  
In this study we consider the evolution of two circumbinary planets that begin on circular orbits that coplanar to each other but misaligned to the binary orbit. Initial planet orbits that are not coplanar to each other are of course possible. For simplicity, we consider
the initially coplanar case for which the initial planet-planet torques are zero. Such a situation could arise
if both planet decouple from the disc as it disperses at about the same time, with a time difference that is short compared to the nodal precession timescale.
We first extend the secular theory for the motion of a single non-zero mass circumbinary planet from  \citet{MartinandLubow2019} to the case of two planets.  We describe the secular evolution of a two planet system around a binary by using analytic models in Section~\ref{ana} for this case. We also extend  the numerical three-body simulations in \citet{Chen20192,Chen20201} to four-body numerical simulations. In Section~\ref{sim}, we describe the initial set-up of our four-body simulations and we describe the results in Section~\ref{mild}. In Section~\ref{crit}, we describe the critical semi-major axes of the outer planet for which the planets undergo mutual libration and circulation.   We discuss the secular nodal resonances between the two planets  in Section~\ref{resonance}. In Section~\ref{high}, we consider planet orbits that are highly inclined with respect to the binary orbital plane. We extend the analytic model to nearly polar orbits and compare the results to numerical simulations. Finally,  we present our discussion and conclusions in Section~\ref{dis}.

\label{int}

\section{Secular Evolution of CIRCUMBINARY PLANET orbits that are nearly coplanar with the binary }
\label{ana}

In this section, we describe an analytic model for the secular evolution of a system of two circumbinary planets that orbit around  a circular or eccentric binary. The planet orbits are initially coplanar  to each other, but slightly misaligned with  respect to the orbit of the binary.
Section 
\ref{high} describes the highly misaligned, nearly polar case.
We
apply the quadrupole approximation for the binary potential based on \citet{Farago2010} and extend the analytic methods of \citet{Lubow2001} and  \citet{Lubow2016}. 
The effects of the planets on the binary orbit are ignored in this model.
We apply a Cartesian coordinate system $(x, y, z)$ in the inertial frame in which the origin is the  binary centre of mass.  The binary initially lies in the $x - y$ plane with its angular momentum vector along the positive $z$ direction. For eccentric orbit binaries, the initial binary eccentricity vector, $\bm e_{\rm b}$, lies along the positive $x$-direction. To study the tilt evolution, we apply the tilt vectors $\bm{\ell}_{\rm p1}=(\ell_{\rm p1x}, \ell_{\rm p1y}, \ell_{\rm p1z})$ and $\bm{\ell}_{\rm p2}=(\ell_{\rm p2x},\ell_{\rm p2y},\ell_{\rm p2z})$ that are angular momentum unit vectors of the inner and outer planets, respectively. We apply the approximation that the two planets have small tilts  with 
respect to the binary orbital plane, $|\ell_{\rm p1x}|\ll 1$,  $|\ell_{\rm p1y}|\ll 1$ and similarly for the outer planet.
We assume that  $\ell_{\rm p1z}\approx \ell_{\rm p2z}
\approx 1$, so that both planets are orbiting in a prograde sense with respect to the binary.
Since the tilt vectors are unit vectors, only the evolution of the $x$ and $y$ components needs to be analysed. We apply linear equations and adopt the time dependence of tilt vectors of the form $\exp(i\omega t)$. Because the components are complex numbers, we take their real parts; for instance, $\ell_{\rm p1x}(t) = {\rm Re}[\ell_{\rm p1x}\exp(i\omega t)]$. There are four eigenmodes, each with a value of $\ell_{\rm p1x}$, $\ell_{\rm p1y}$, $\ell_{\rm p2x}$, $\ell_{\rm p2y}$, and $\omega$. 

The interaction between two planets which have masses $m_{\rm p1}$ and $m_{\rm p2}$ is described by a coupling coefficient denoted by 
\begin{equation}
C_{\rm p1,p2} = G m_{\rm p1} m_{\rm p2}K(r_{\rm p1}, r_{\rm p2}), 
\label{eq:pp}
\end{equation}
where the symmetric kernel that has units of inverse length is 
\begin{equation}
\begin{aligned}
K(r_{\rm p1}, r_{\rm p2}) = \frac{r_{\rm p1} r_{\rm p2}}{4\pi} \int_{0}^{2\pi}\frac{\cos\phi \, d\phi}{(r_{\rm p1}^2+r_{\rm p2}^2-2r_{\rm p1} r_{\rm p2} \cos\phi)^{3/2}},
\end{aligned}
\end{equation}
where $r_{\rm p1}$ and $r_{\rm p2}$ are distances of each planet to the initial centre of the mass of the binary.
The simplified analytic form of the kernel is
\begin{equation}
\begin{aligned}
K(r_{\rm p1}, r_{\rm p2}) = \frac{\left(\rm{r_1}^2+\rm{r_2}^2\right) E_2\left(\frac{4 \rm{r_1}  \rm{r_2}}{(\rm{r_1}+\rm{r_2})^2}\right)-(\rm{r_1}-\rm{r_2})^2 \it{E_1} \left(\frac{4 \rm{r_1} \rm{r_2}}{(\rm{r_1}+\rm{r_2})^2}\right)}{2 \pi  (\rm{r_1}-\rm{r_2})^2 (\rm{r_1}+\rm{r_2})},
\end{aligned}
\end{equation}
where $E_2$ is the complete elliptic integral of the second kind and $E_1$ is the complete elliptic integral of the first kind.

The nodal precession frequency of a single planet on a nearly coplanar orbit about a circular orbit binary is approximately given by 
\begin{equation}
\omega_{\rm c}(r) = \frac{3}{4}\frac{M_1 M_2}{M_{\rm b}^2}\Omega_{\rm b}\left(\frac{a_{\rm b}}{r}\right)^{7/2},
\end{equation}
where $a_{\rm b}$ is the semi-major axis of the binary which has components
of mass $M_1$, $M_2$ with total mass $M_{\rm b} = M_1 + M_2$,
and  $\Omega_{\rm b}$ is the the binary orbital frequency. The single-planet nodal precession frequency associated with each slightly inclined planet is 
\begin{gather}
    \omega_{\rm p1} = \omega_{\rm c}(r_{\rm p1}), \label{omegap1} \\
    \omega_{\rm p2} = \omega_{\rm c}(r_{\rm p2})\label{omegap2}.
\end{gather}
The evolution equations for planet tilts  $ \ell_{\rm p1}(t)$ and $ \ell_{\rm p2}(t)$ are given by
\begin{gather}
    i \omega J_{\rm p1} \ell_{\rm p1x} = C_{\rm p1,p2} (\ell_{\rm p1y}- \ell_{\rm p2y}) + \tau_{\rm x} \omega_{\rm p1} J_{\rm p1}  \ell_{\rm p1y},\label{i1}\\     
    i  \omega J_{\rm p1} \ell_{\rm p1y} = C_{\rm p1,p2}(\ell_{\rm p2x}- \ell_{\rm p1x}) + \tau_{\rm y} \omega_{\rm p1} J_{\rm p1}  \ell_{\rm p1x},\\     
    i   \omega J_{\rm p2} \ell_{\rm p2x} = C_{\rm p1,p2}(\ell_{\rm p2y}- \ell_{\rm p1y}) + \tau_{\rm x} \omega_{\rm p2} J_{\rm p2}  \ell_{\rm p2y},\\     
    i  \omega J_{\rm p2} \ell_{\rm p2y} = C_{\rm p1,p2}(\ell_{\rm p1x}- \ell_{\rm p2x}) + \tau_{\rm y} \omega_{\rm p2} J_{\rm p2}  \ell_{\rm p2x}\label{i4},     
\end{gather}
where the angular momenta of the planets that orbit at radii $r_{\rm p1}$ and $r_{\rm p2}$ are
(approximately) given by $J_{\rm p1}=m_{\rm p1} r_{\rm p1}^2 \Omega_{\rm k}(r_{\rm p1})$  and $J_{\rm p2}=m_{\rm p2} r_{\rm p2}^2 \Omega_{\rm k}(r_{\rm p2})$, respectively, for Keplerian orbital frequency around the binary $\Omega_{\rm k}(r)=\sqrt{GM_{\rm b}/r^3}$. Moreover, $\tau_{\rm x}$ and $\tau_{\rm y}$ are related to the secular torque on both planets due to the binary with the eccentricity $e_{\rm b}$. We apply torque equations 2.16-2.18 in \citet{Farago2010} to the nearly coplanar case and obtain
\begin{equation}
    \tau_{\rm x} = (1-e_{\rm b}^2)
\end{equation}
    and
\begin{equation}
    \tau_{\rm y} = -(1+4e_{\rm b}^2). \label{tauyp}
\end{equation}

We solve these equations analytically in Mathematica. First, we normalise $\ell_{\rm p1x}$ = 1 so that Equations (\ref{i1}) - ~(\ref{i4}) are solved  for  $\ell_{\rm p1y}$,  $\ell_{\rm p2x}$,  $\ell_{\rm p2y}$, and $\omega$ in terms of $C_{\rm p1, p2}$, $J_{\rm p1}$, $J_{\rm p2}$, and $e_{\rm b}$. The solution provides four eigenmodes and eigenfrequencies. Second, we impose the initial conditions to determine the weights of the modes that are applied to derive the unique solution. Each mode is represented as a column in a 4x4 matrix $\cal M$ in the form of the modal solutions
for ($\ell_{\rm p1x}$, $\ell_{\rm p1y}$, $\ell_{\rm p2x}$, $\ell_{\rm p2y}$)$^{\rm T}$, where $\rm T$ represents the transpose operator that transforms the row vector into a column vector. The weights $w_{j}$ for mode number $j$ = 1, 2, 3, 4 are represented in column vector $\bm{w}$ that is computed by 
\begin{equation}
    \bm{w} = {\cal M}^{-1}{\bm{\ell}_0}, 
\end{equation}
where $\bm{\ell}_0$ is the column vector that contains the known initial tilts ($\ell_{\rm p1x}$, $\ell_{\rm p1y}$, $\ell_{\rm p2x}$, $\ell_{\rm p2y}$)$^{\rm T}$. The weight column vector $\bm{w}$ is determined analytically through the matrix inversion. The full solution for the tilts in time is then given by 
\begin{equation}
   \bm{ \ell}(t) = {\rm Re}[{\cal M}\bm{w}(t)],
\end{equation}
where $\bm{\ell} (t)$ is the solution column vector ($\ell_{\rm p1x}(t)$, $\ell_{\rm p1y}(t)$, $\ell_{\rm p2x}(t)$, $\ell_{\rm p2y}(t)$)$^{\rm T}$ and $\bm{w}(t)$ is the column vector ($w_1$exp($i\omega_1 t$),...,$w_4$exp($i\omega_4 t$))$^{\rm T}$ with subscripts denoting the mode number. 

In evaluating the results, we calculate each planet's inclination in the small angle approximations with respect to the initial binary orbital plane as
\begin{equation}
    i_{\rm pj}(t) = \sqrt{\ell_{\rm pjx}^2(t)+\ell_{\rm pjy}^2(t)}\ i_0,
\label{eq:ip}
\end{equation}
where $j$ = 1, 2 and $i_0$ is the initial tilt value. We determine the longitude of ascending node of each planet as 
\begin{equation}
    \phi_{\rm pj}(t) = \tan^{-1}\left( \frac{-\ell_{\rm pjx}(t)}{\ell_{\rm pjy}(t)}\right),
\label{eq:phi}
\end{equation}
where $j$ = 1, 2. We determine the relative inclination between the two planets that begin at a common inclination $i_{\rm 0}$ as
\begin{equation}
    \Delta i_{\rm p}(t) =  \sqrt{(\ell_{\rm p1x}(t)-\ell_{\rm p2x}(t))^2+(\ell_{\rm p1y}(t)-\ell_{\rm p2y}(t))^2} \ i_0. \label{Di}
\end{equation}
In Section~\ref{mild} we show some analytic results for the nearly coplanar case and compare them to numerical four-body simulations.

\section{Four-body simulations of CIRCUMBINARY PLANET orbits that are nearly coplanar with the binary}
\label{sim}

To study the orbital dynamics of two massive bodies orbiting around a binary system, we make use of the ${\sc n}$-body simulation code, {\sc rebound} and apply the {\sc whfast} integrator. This is a second order symplectic Wisdom-Holman integrator with 11th order symplectic correctors \citep{Rein2015b}. The equations of motion for the four bodies are solved in the inertial frame with the origin at the centre of mass of the four-body system. We consider an equal mass binary. 


 The initial conditions for the two planets are those of Keplerian orbits
about a point mass equal to the binary mass. Both of planets have the same mass $m_{\rm p1} = m_{\rm p2} = 0.001\,M_{\rm b}$.
Their orbits are defined by six orbital elements: the semi-major axes $a_{\rm p1}$ and $a_{\rm p2}$, inclinations $i_{\rm p1}$ and $i_{\rm p2}$ relative to the binary orbital plane, eccentricities $e_{\rm p1}$ and $e_{\rm p2}$, longitudes of the ascending node $\phi_{\rm p1}$ and $\phi_{\rm p2}$, arguments of periapsis $\omega_{\rm p1}$ and $\omega_{\rm p2}$, and true anomalies $\nu_{\rm p1}$ and $\nu_{\rm p2}$. The orbits of the two planets are initially circular so that $e_{\rm p1}=e_{\rm p2}=0$ and initially $\omega_{\rm p1}=\omega_{\rm p2}=0$ and $\nu_{\rm p1}=\nu_{\rm p2}$ = 0. We choose $\phi_{\rm p1}$ = $\phi_{\rm p2}=90^{\circ}$  initially in our suites of simulations. Table~\ref{table1} shows
the initial values of $e_{\rm b}$, $a_{\rm p1}$, $a_{\rm p2}$, $i_{0}$ that we apply. 
The binary orbit is not fixed in these simulations. The binary can evolve due to the gravity of the massive third body ($p1$) and fourth body ($p2$). 
For comparison to the analytic solutions, we determine the inclination of each planet $j$ by
\begin{equation}
i_{\rm pj}(t) = \arccos{({\ell_{\rm pjz}}(t))} \label{ipn}
\end{equation}
and use Equation (\ref{eq:phi}) for the longitude of the ascending node of each planet in simulations. 
In the next section we discuss the results of these four body simulations with  low inclination orbits and  compare them with results from the analytic model of Section \ref{ana}.

\section{Results for CIRCUMBINARY PLANET orbits that are nearly coplanar with the binary}
\label{mild}

\subsection{Circular orbit binaries with two nearly coplanar planets with respect to the binary}
\label{sec:tilt}

\subsubsection{Fiducial model}
\label{sec:fudicial}

\begin{table}
\centering
\caption{Parameters of the simulations. Column 1 is the name of the simulation. Column 2 is the initial eccentricity of the binary. Column 3 is the initial semi-major axis of the inner planet. Column 4 is the the initial semi-major axis of the outer planet. Column 5 is the initial inclination of both planets relative to the binary orbit that is taken to be the same for both planets. Column 6 is the simulation end time. }
\label{tab:example_table}
\begin{tabular}{cccccc} 
\hline
\textbf{Model} & $e_{\rm b}$ & $a_{\rm p1}$ & $a_{\rm p2}$ & $i_0$ &Simulation time \\
 & & $(a_{\rm b})$& ($a_{\rm b})$& (deg) & ($10^{6}\times T_{\rm b}$)\\
\hline
A0 & 0.0  & 10 & 14.5 &1, 10 & 5\\
B0 & 0.0  & 10 & 19.0 &1, 10 & 5\\
A1 & 0.2  & 10 & 14.5 &1, 10 & 5\\
A2 & 0.5  & 10 & 14.5 &1, 10 & 5\\
A3 & 0.8  & 10 & 14.5 &1, 10 & 5\\
C1 & 0.0 & 5 & 9.5& 1, 10 & 1\\
C2 & 0.0 & 20 & 29.0& 1, 10 & 10\\
T1 & 0.8 & 11.4 & 17.0& - & 0.2\\
T2 & 0.8 & 12 & 16.02& - & 0.2\\
T3 & 0.8 & 12 & 21.18& - & 0.2\\
H1 & 0.5 & 10.0 & 18.0& 80 & 4\\
H2 & 0.5 & 10.0 & 18.0& 90 & 4\\
\hline
\label{table1}
\end{tabular}
\end{table}

\begin{figure*}
\includegraphics[width=18cm]{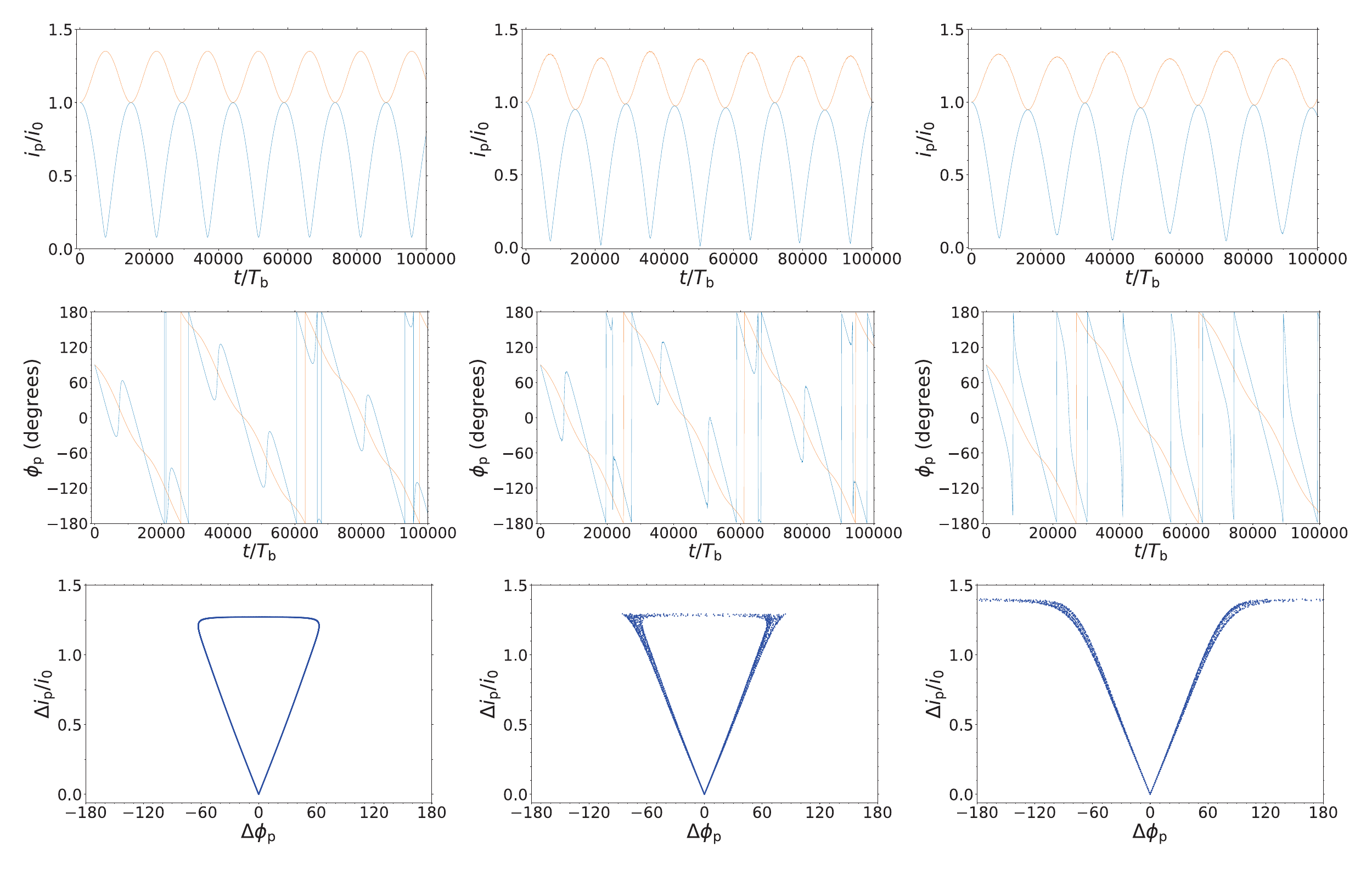}
\caption{Model A0. Time evolution of two circumbinary planets that orbit around the binary with the parameters described in Table~\ref{table1} (model A0). Left panels: analytic model, middle and right panels: numerical simulations with $i_{0}=1^\circ$ (middle) and  $i_{0}=10^\circ$ (right). The blue lines represent the inner planet and the yellow lines represent the outer planet. The upper panels show the planet inclinations relative to the binary orbital plane, the middle panels show the nodal phase angles of the planets, and the lower panels show the relative tilt between two planets as a function of their nodal phase angle difference. }
\label{aA0}
\end{figure*}

\begin{figure*}
\includegraphics[width=18cm]{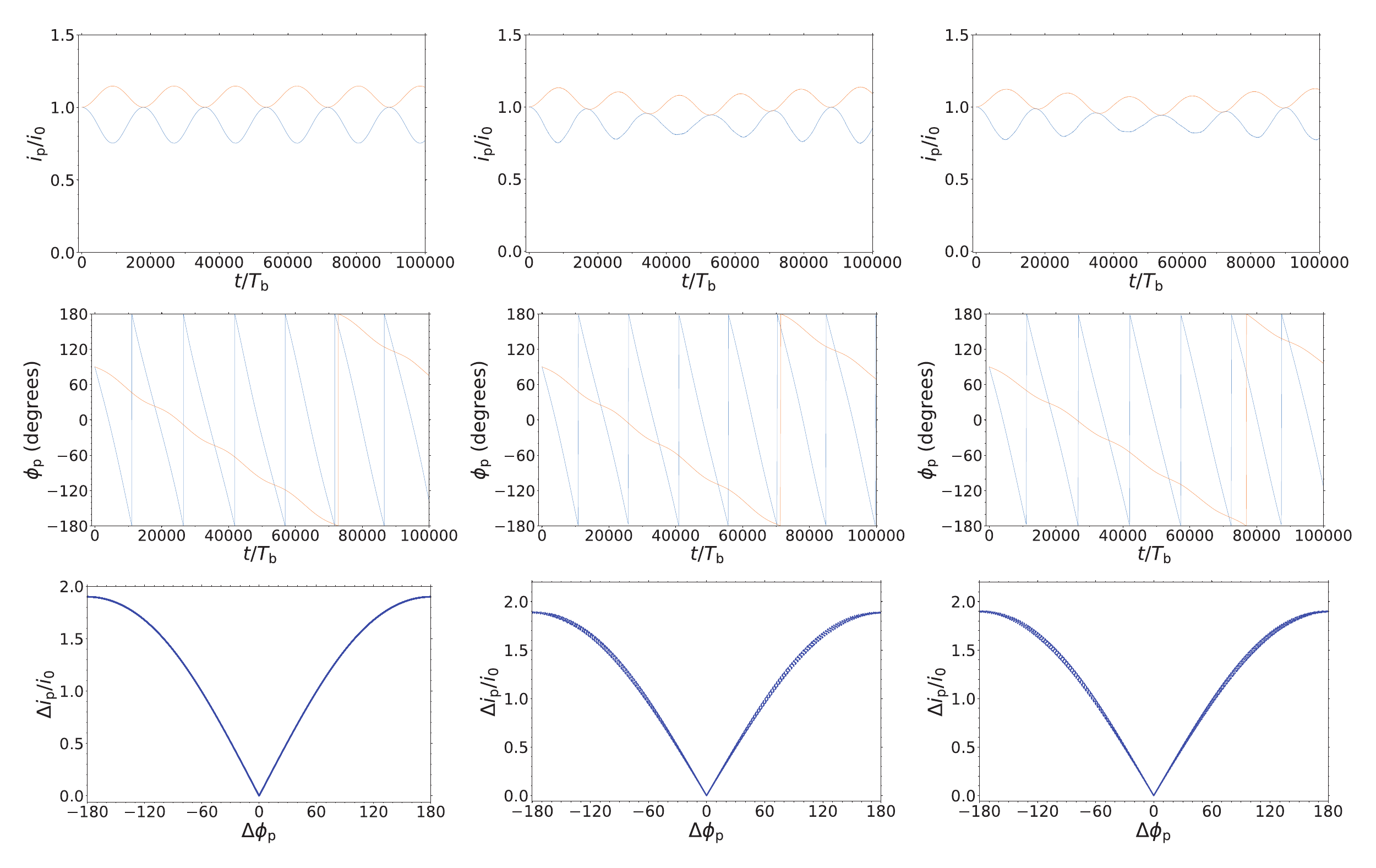}
\caption{Model B0. Same as Model A0 in Figure~\ref{aA0} except $a_{\rm p2}$ = 19.0 $a_{\rm b}$.}
\label{aB0}
\end{figure*}

\begin{figure*}
\includegraphics[width=18cm]{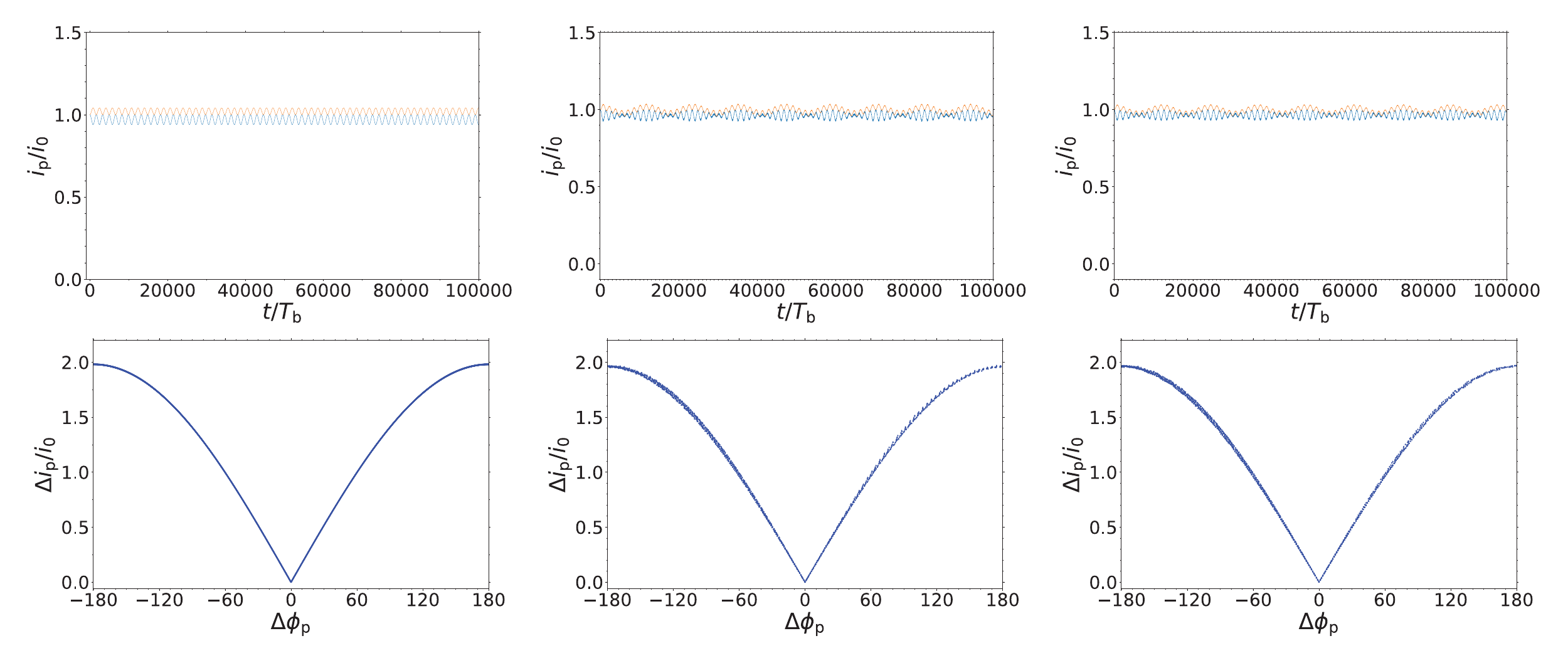}
\caption{Model C1. Same as Model A0 in Figure~\ref{aA0} except $a_{\rm p1}$ = 5.0$a_{\rm b}$ and $a_{\rm p2}$ = 9.5 $a_{\rm b}$} 
\label{aC1}
\end{figure*}

\begin{figure*}
\includegraphics[width=18cm]{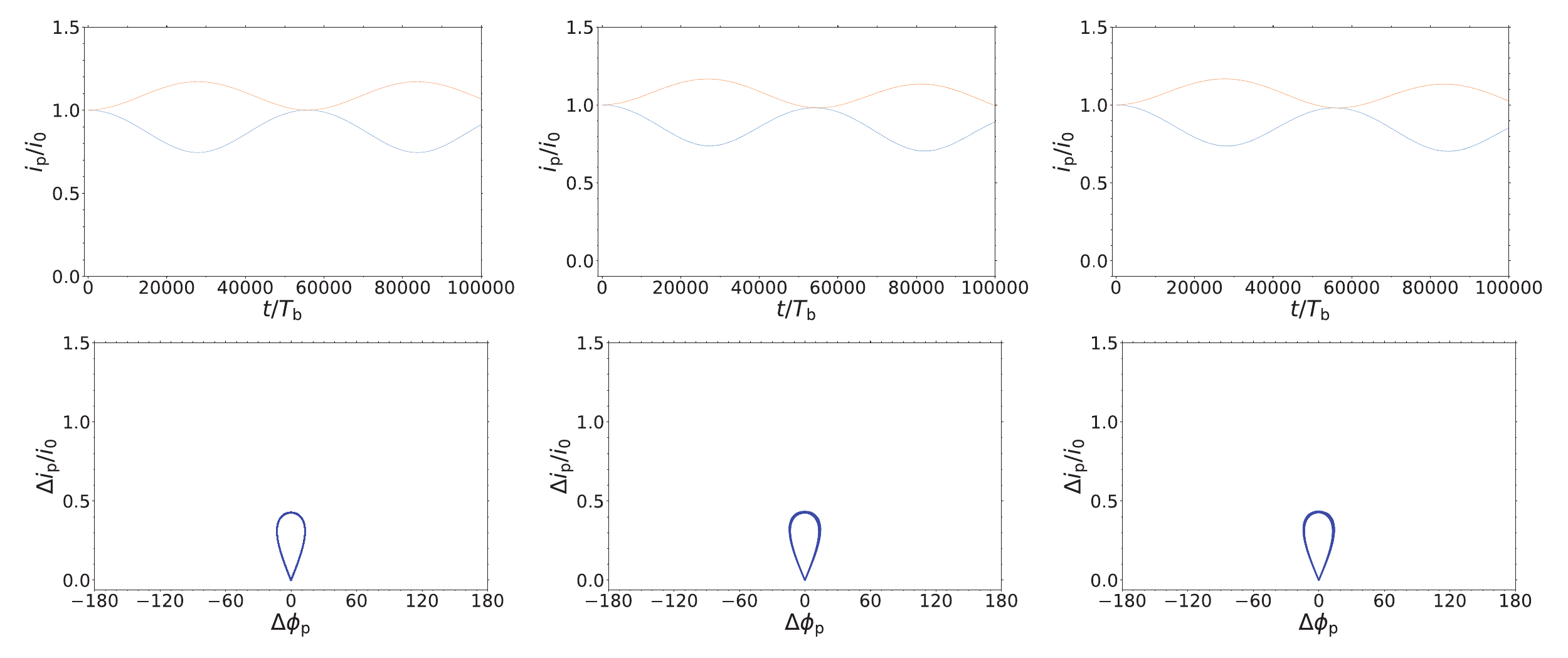}
\caption{Model C2. Same as Model A0 in Figure~\ref{aA0} except $a_{\rm p1}$ = 20.0 $a_{\rm b}$ and $a_{\rm p2}$ = 24.5 $a_{\rm b}$} 
\label{aC2}
\end{figure*}

To first understand the interaction between two circumbinary planets, we consider a fiducial model (model A0) in which the binary has eccentricity $e_{\rm b} = 0$. The inner planet is at semi-major axis $a_{\rm p1}=10\,a_{\rm b}$ and the outer planet is at $a_{\rm p2}=14.5 \,a_{\rm b}$. The two planets have the same initial inclination $i_{\rm p} = i_0$.  Figure~\ref{aA0} shows the time evolution of $i_{\rm p}$ (upper panels), $\phi_{\rm p}$ (middle panels) and the phase plots of the relative tilt between two planets as a function of their nodal phase difference (lower panels). The left panels show the analytic model while the middle and  right panels show the numerical models with $i_0 = 1^\circ$ and $i_0 = 10^\circ$, respectively. 

The blue lines in Figure~\ref{aA0} represent the inner planet while the yellow lines represent the outer planet. During the tilt oscillations, the inclination of the inner planet $i_{\rm p1}$ initially drops to $0.05 i_0$ while the inclination of the outer planet $i_{\rm p2}$ increases to $1.35 i_0$. The interaction between the  two planets also affects the nodal precession angle $\phi_{\rm p}$, as shown in the middle row. Without the inner planet, the outer planet would precess more slowly than the inner planet because it is farther away from the binary. However, the analytic solution (left) and the low inclination simulation (middle) show the two planets locked to each other and evolving on the same average precession timescale due to the planet-planet interactions. This kind of behaviour can be explained by the lower panels which  show  the phase of the relative tilt between two planets as a function of their nodal phase difference. As the phase angles of two planets are coupled together by their mutual gravitational interaction, the nodal phase difference is limited in range and oscillates about zero. The two planets are then mutually librating. 

Comparing the analytic model (left) with numerical model with $i_0=1^\circ$ (middle), we see that there is good agreement.
In the four body simulation, the interaction between the binary and the planets results in the precession of the angular momentum and eccentricity vectors of the binary. The binary precession in turn affects
the orbital evolution of the planets. This effect is small and not
included in the analytic model of Section \ref{ana}.

However, in comparing to the numerical model with $i_0=10^\circ$ (right), we find the oscillation timescale of the higher inclination simulation is slightly longer than the other two models. In the phase plots, the precession rates of the inner planet and the outer planet are different in the simulation with $i_0=10^\circ$ as the two planets are not locked to each other. Thus, in the lower-right panel, the nodal phase difference varies from 0 to 180$^\circ$. The two planets are then mutually circulating.

\subsubsection{Effect of the planet separation}

The strength of the interaction between the two planets decreases with increasing separation between them. Figure~\ref{aB0} shows a simulation with the same parameters as model A0, except that $a_{\rm p2}$ is larger at 19$\,a_{\rm b}$. The orbital separation between the two planets is two times larger than model A0. In this case, the analytic model (left column) is in good agreement with the numerical simulations at $i_0=1^\circ$ (middle column),  although we can see some differences in the upper panels due to the precession of the angular momentum vector of binary. The upper panels show that the inclination of the inner planet, $i_{\rm p1}$, only drops to 0.75$\, i_0$ while the inclination of the outer planet, $i_{\rm p2}$, increases up to 1.3$\,i_0$ during the tilt oscillations. The nodal phase angle of the outer planet, $\phi_{\rm p2}$ (shown in the middle row panels) has a longer precession timescale than the inner planet, $\phi_{\rm p1}$, since the planets are not locked to each other, even though two planets have tilt oscillations. The lower panels show that the system undergoes circulation in all models.

\subsubsection{Effect of the semi-major axis of the planets}

We now consider how the semi-major axes of the inner planet $a_{\rm p1}$ and the outer planet $a_{\rm p2}$ affect the evolution of the two planets. We first consider the two planets being closer to the binary and then the two planets being farther away than in our fiducial model A0. 

Figure~\ref{aC1} shows the evolution of model C1 which has the same parameters as the fiducial model A0, except that both planets are closer to the binary with  $a_{\rm p1} = 5 \,a_{\rm b}$ and $a_{\rm p2} = 9.5 \,a_{\rm b}$.  The upper panels show that there are still tilt oscillations, but $i_{\rm p2}$ only increases up to 1.02$\,i_0$, while $i_{\rm p1}$ only decreases to 0.95$\,i_0$ even though the separation between two planets is similar to model A0.  Comparing with the simulations in middle and right panels, we see that all of them have similar maximum and minimum amplitudes. The tilt oscillation period is much shorter than model A0 because the planets are closer to the binary. The precession rate of the angular momentum vector of the binary is faster than in model A0. 
and so wave packets can be seen in simulation panels. 
The lower panels show that the nodal phase angles of the two planets are not  locked to each other so the system undergoes circulation for both the analytic and numerical models.  

Figure~\ref{aC2} shows the evolution of model C2 which has the same parameters as model A0 except the planets are both farther away from the binary with $a_{\rm p1} = 20.0 \,a_{\rm b}$ and $a_{\rm p2} = 29 \,a_{\rm b}$.  Again, the analytic model and the numerical simulations are very similar. In addition to the longer tilt oscillation period than model A0, the system has a longer precession timescale of the angular momentum vector of the binary due to the larger separation between the binary and planets. Thus, the effect of the precession of the angular momentum of the binary is not so obvious within only 100000$\,T_{\rm b}$.  The inclinations of the two planets initially oscillate away from each other and  $i_{\rm p2}$ increases to 1.2 $i_0$ while $i_{\rm p1}$ drops to 0.7 $i_0$. In the lower panel, unlike model~B0,  the two planets are locked to each other in the same average nodal phase angle and the system undergoes libration. 

Because the stronger effect of the binary in Model C1 causes the planets to precess more rapidly at different frequencies, they  unable to precess together due to their mutual gravitational interaction. Because the weaker effect of the binary in Model C2 causes the planets to precess
more slowly, they are able to precess together.

\subsection{Eccentric orbit binaries with two nearly coplanar planets with respect to  the binary}

We now consider how the binary eccentricity, $e_{\rm b}$, affects the planet-planet interactions. Even in the case of a single planet, the eccentric binary orbit leads to tilt oscillations of the  planet because it produces a nonaxisymmetric
secular potential \citep{Farago2010,Smallwood2019}.  

\begin{figure*}
\includegraphics[width=18cm]{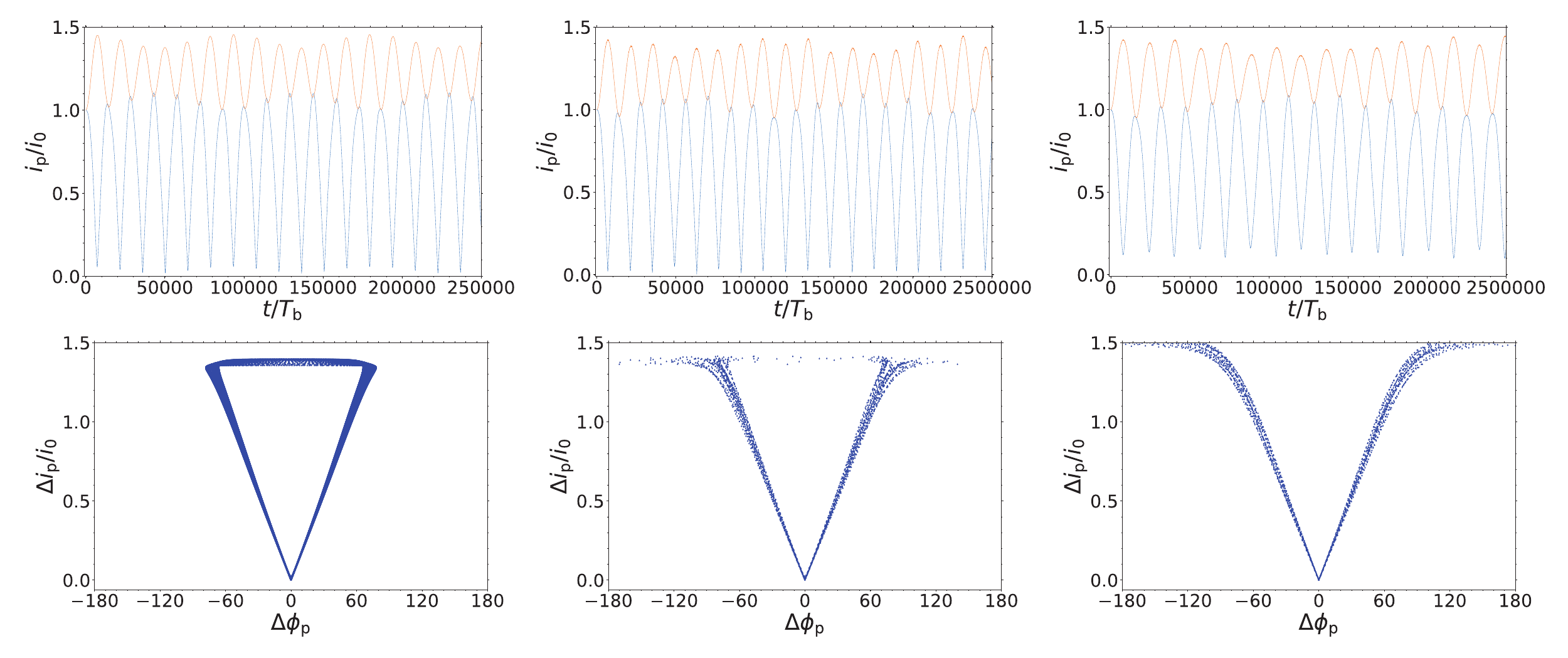}
\caption{Model A1. Same as Model A0 in Figure~\ref{aA0} except $e_{\rm b}$ = 0.2.} 
\label{aA1}
\end{figure*}

Figure~\ref{aA1} shows model A1 that has the same parameters as the fiducial model A0, except that the binary eccentricity is $e_{\rm b} = 0.2$. The analytic results agree well with the numerical models.  The maximum inclination of the outer planets is about 1.45$\,i_0$ while the minimum inclination of the inner planet is 0.03$\,i_0$ during the oscillations. 
The two planets are locked 
to each other in the analytic model and are often locked in the simulation with $i_0=1^\circ$. They  are not locked to each other in the simulation with $i_0=10^\circ$. Thus, the analytic model and the simulation with $i_0=1^\circ$ typically undergo libration and the simulation with $i_0=10^\circ$ undergoes circulation.

\begin{figure*}
\includegraphics[width=18cm]{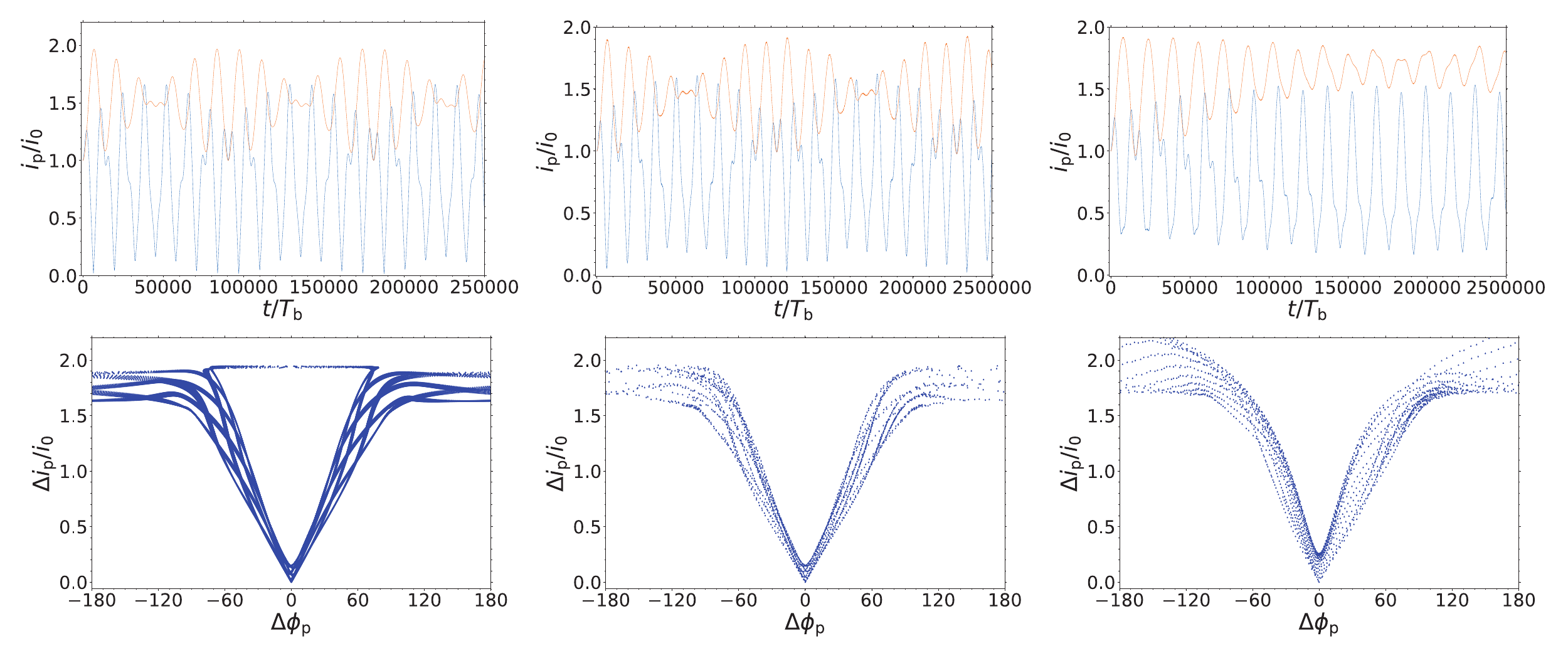}
\caption{Model A2. Same as Model A0 in Figure~\ref{aA0} except $e_{\rm b}$ = 0.5.} 
\label{aA2}
\end{figure*}

Figure~\ref{aA2} shows model A2 that has the same parameters as model A0 except the binary eccentricity is $e_{\rm b} = 0.5$. The planet-binary tilt oscillations are stronger than in model~A1. The inclination of the inner planet, $i_{\rm p1}$, increases up to 2$\,i_0$ while the inclination of the outer planet, $i_{\rm p2}$, decreases to less than 0.1 during the oscillations. 
The analytic model generally undergoes circulation, while the simulation with $i_0 =1^\circ$ always undergoes circulation. All three models are quite similar
but evolve on different timescales. The effect of the higher binary eccentricity than in model~A1 is to cause the planets to unlock and circulate.

\begin{figure*}
\includegraphics[width=18cm]{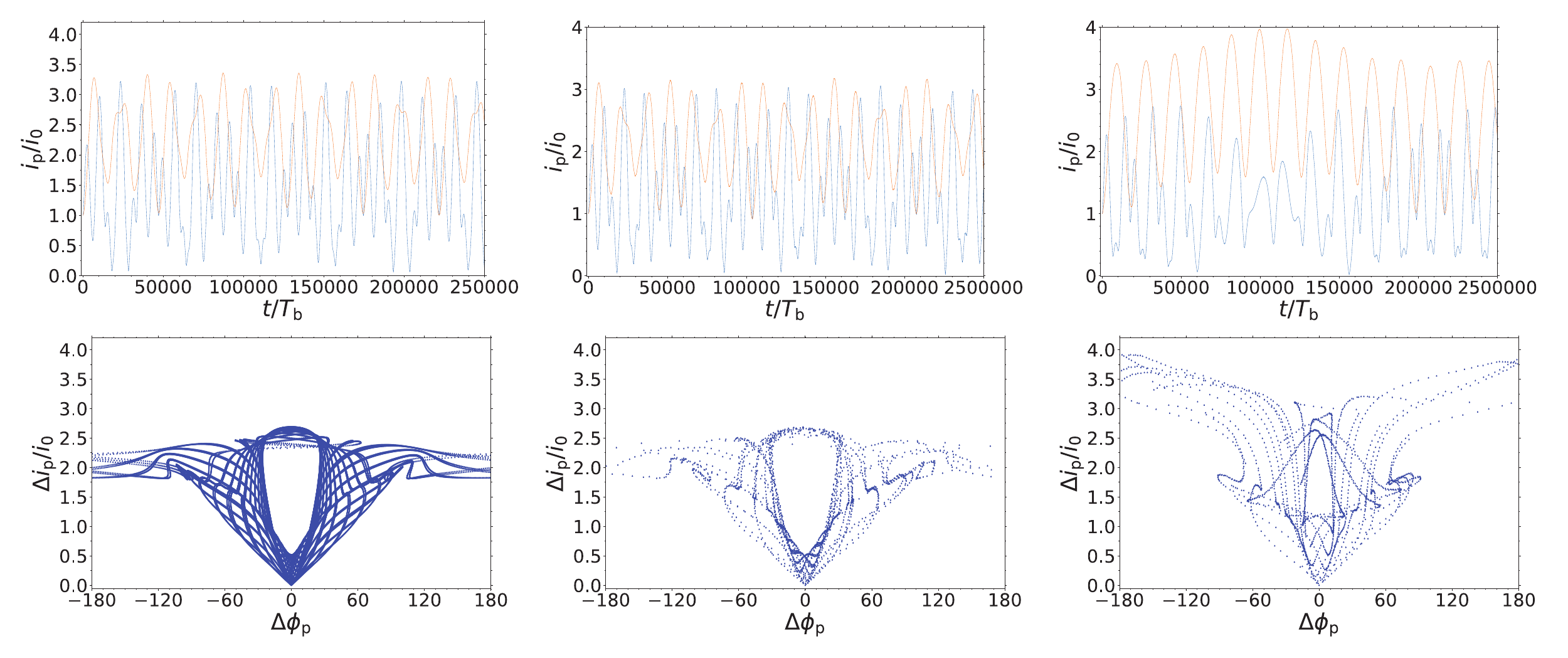}
\caption{Model A3. Same as Model A0 in Figure~\ref{aA0} except $e_{\rm b}$ = 0.8.} 
\label{aA3}
\end{figure*}

Figure~\ref{aA3} shows model A3 that has the same parameters as model A0 except an even higher binary eccentricity of $e_{\rm b} = 0.8$. In the upper panels, we see that the planet-binary tilt oscillations dominate the system. The inner planet inclination $i_{\rm p1}$ increases up to 3.2$\,i_0$ while the outer planet inclination $i_{\rm p2}$ decreases to less than 0.1$\,i_0$ during the oscillations. For the higher initial tilt simulation with $i_0=10^\circ$, the inclination of the inner planet, $i_{\rm p1}$, even increases up to 4.0$\,i_0$
(upper right panel). The two planets are locked and unlocked in the nodal phase angle over time, the system undergoes both circulation and libration in the lower panels and the simulation with $i_0=10^\circ$ has a larger variation than the other two models.

\section{Critical semi-major axis for libration/circulation }
\label{crit}

\begin{figure*}
\includegraphics[width=8.6cm]{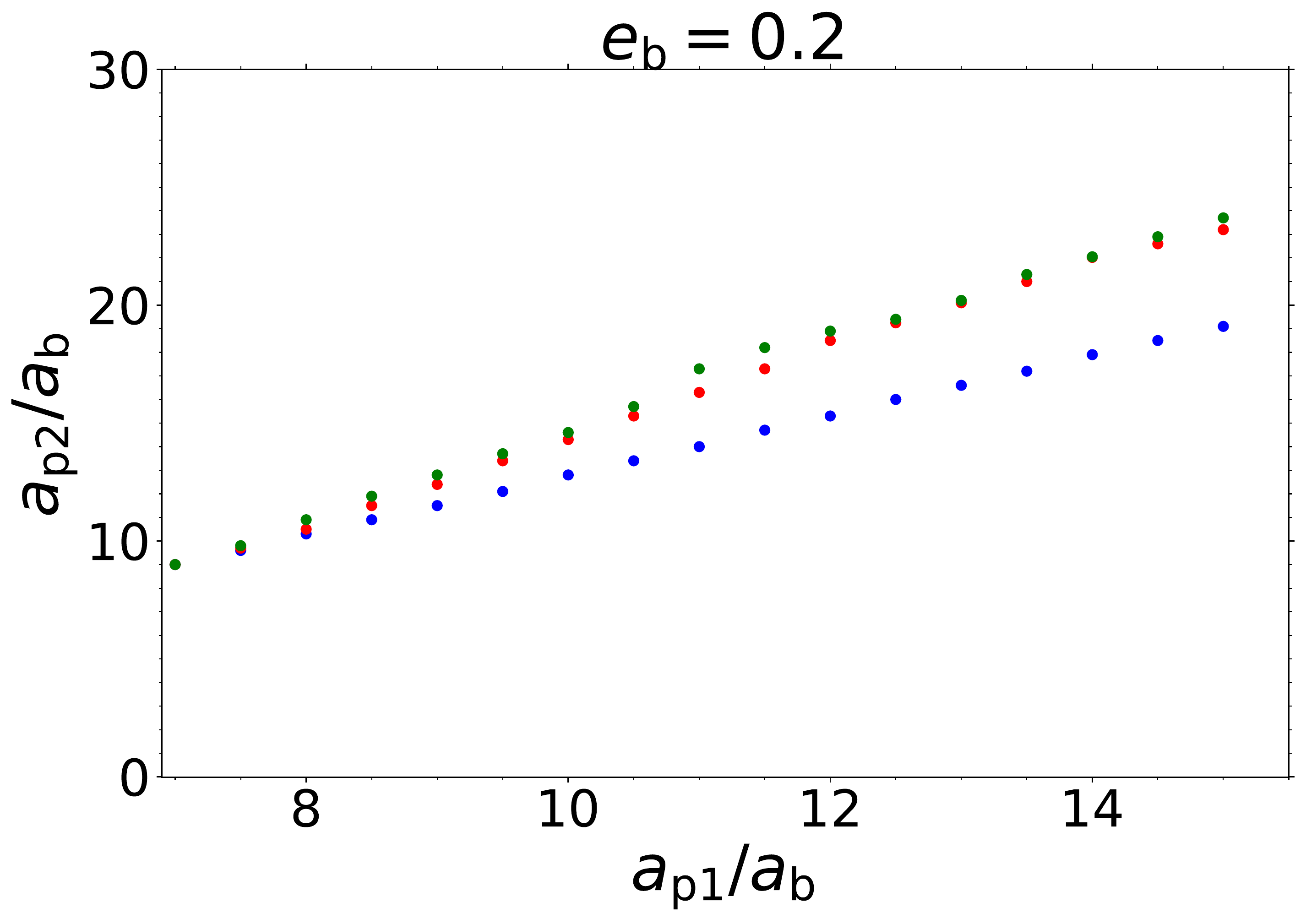}
\includegraphics[width=8.6cm]{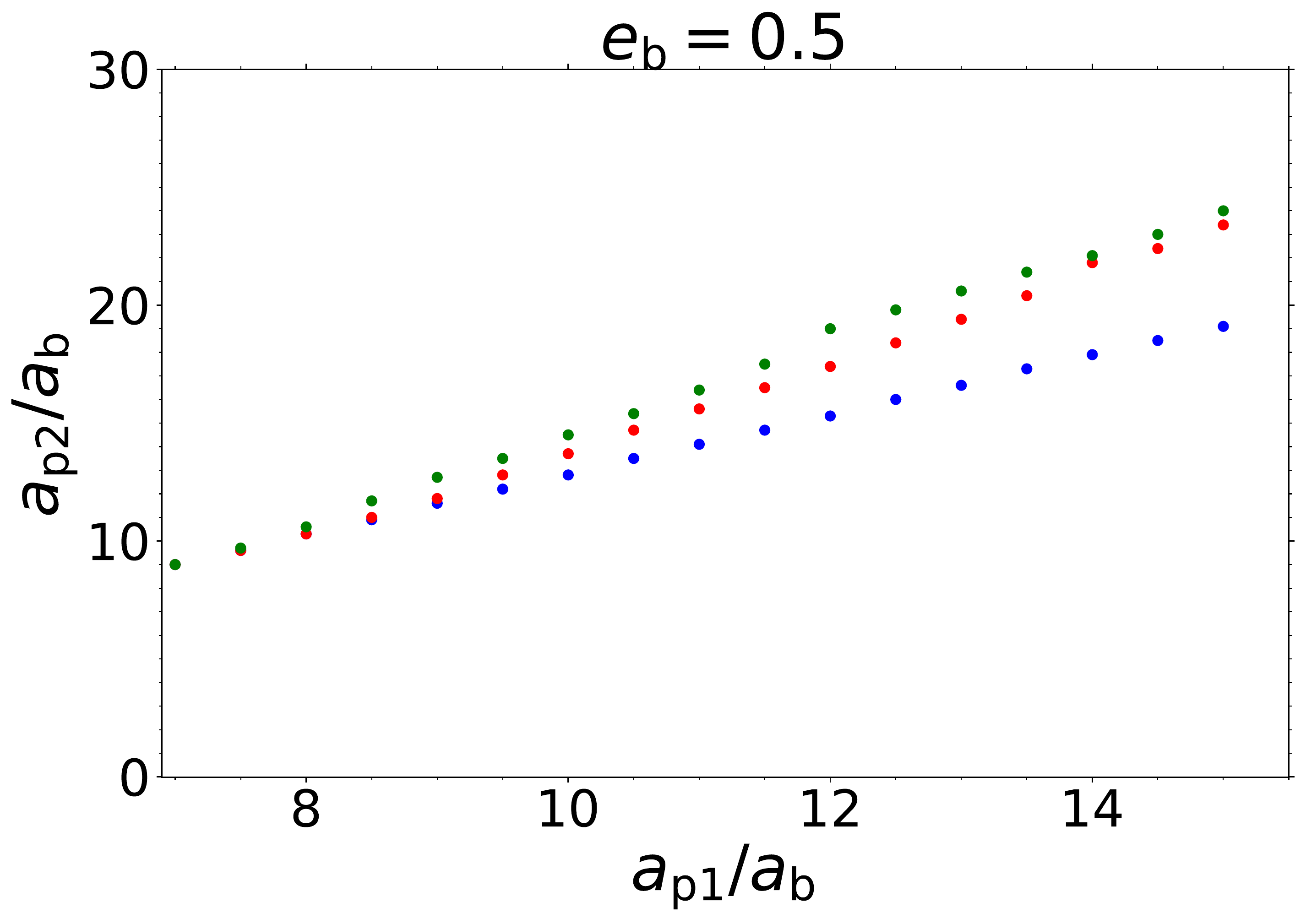}
\includegraphics[width=8.6cm]{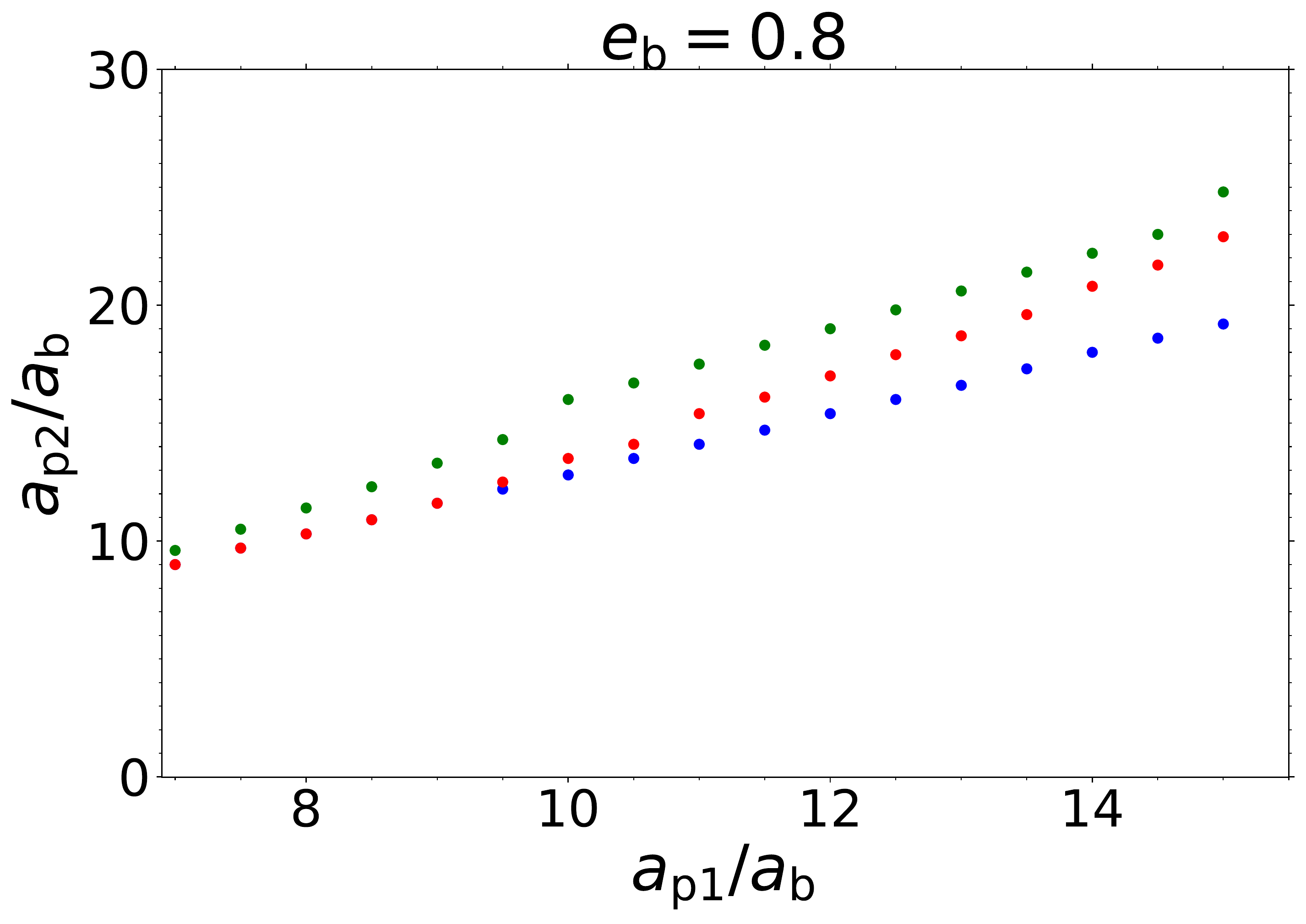}

\caption{Critical semi-major axes of the outer planet for the different orbit types around  binary systems with $e_{\rm b}$ = 0.2 (upper-left), 0.5 (upper-right) and 0.8 (lower). Between the blue and red dots, the system undergoes libration.  Between the red and green dots the system undergoes both libration and circulation. Above the green dots the system is completely circulating. The system has $a_{\rm p1}$ ranging from 7 to 15 $a_{\rm b}$ with interval 0.5 $a_{\rm b}$. The two planets have $i_0$ =1$^{\circ}$.} 
\label{cri}
\end{figure*}

We consider planets that are nearly coplanar with respect to the binary.
We investigate the conditions under which the planets undergo mutual libration, in which the difference in the longitudes of ascending node is limited to be less than $360^\circ$,
and mutual circulation, in which the difference in the longitudes of ascending node  reaches $360^\circ$.
As the separation between two planets increases, the interaction between two circumbinary planets changes from libration to circulation with respect to each other. However, there is a range of radii for which the system undergoes both circulation and libration before it becomes completely circulating, as seen in  Figure~\ref{aA3}. To better understand the transition between these two behaviours with different $e_{\rm b}$, in Figure~\ref{cri}, we plot the critical semi-major axes of the outer planet as a function of the semi-major axis of the inner planet. We consider three different binary eccentricities $e_{\rm b}$ = 0.2 (upper-left), 0.5 (upper-right) and 0.8 (lower) with different initial semi-major axes of the inner planet ranging from 7$\,a_{\rm b}$ to 15$\,a_{\rm b}$ with an interval of 0.5$\,a_{\rm b}$. The blue dots represent the critical semi-major axes of the outer planet beyond which the two planets are able to undergo libration. Between the blue and red dots, the system is always librating. Between the red and green dots, the system is both circulating and librating. Above the green dots, the system is always circulating. The size of the region where the system is both circulating and librating increases with increasing binary eccentricity. 

We explain these results for planet orbits that are nearly coplanar with
respect to the binary orbital plane in terms of the analytic model. For a circular orbit binary,
there is a sharp transition from mutual libration to circulation. That this, 
there are no intermediate orbits that show properties of both libration and circulation.
The tilt oscillations in the circular orbit binary case are simple. The 
oscillations are due to planet-planet
interactions only, since the secular potential of the binary is axisymmetric and cannot drive a tilt change. The tilt of each planet depends only on the nodal phase difference between the planets. The tilts are therefore periodic with the same frequency for both planets, as seen in Figures~\ref{aA0} to \ref{aC2}. 
 An intermediate orbit that involves both libration and
circulation is not periodic and so  cannot occur around a circular orbit binary.

An eccentric orbit binary provides a nonaxisymmetric secular potential.
As a result, the eccentric orbit binary causes additional tilt frequencies to be present. In general, four different tilt frequencies, involving two independent frequencies,  occur in the analytic
secular model for an eccentric orbit binary. (The other two frequencies
differ in sign from these frequencies.) As a result, complex nonperiodic orbits generally arise that
allow for the existence of these intermediate orbits that undergo
a combination of libration and circulation, as we find in Figure~\ref{aA3}. Based on this argument, we would
expect the range of parameters over which these intermediate orbits exist to
increase with binary eccentricity, as we find in Figure~\ref{cri}.

\section{Secular nodal resonance between two planets}
\label{resonance}

When we solve Equations (\ref{i1}) - ~(\ref{i4}) analytically,  the solution includes four eigenmodes and hence has four values for the eigenfrequency $\omega$. Of these four eigenfrequencies, there are two  positive values that we denote with $\omega_1$ and  $\omega_2$. (The other two values are $-\omega_1$ and $-\omega_2$.)
A secular resonance occurs when the ratio $\omega_1$/$\omega_2$ is simple integer ratio. When the system is in resonance, the complex pattern in the phase diagram repeats itself exactly. This phenomenon can be seen in systems undergoing libration, as well as systems undergoing circulation, and cases that display both behaviours.

\begin{figure*}
\includegraphics[width=18cm]{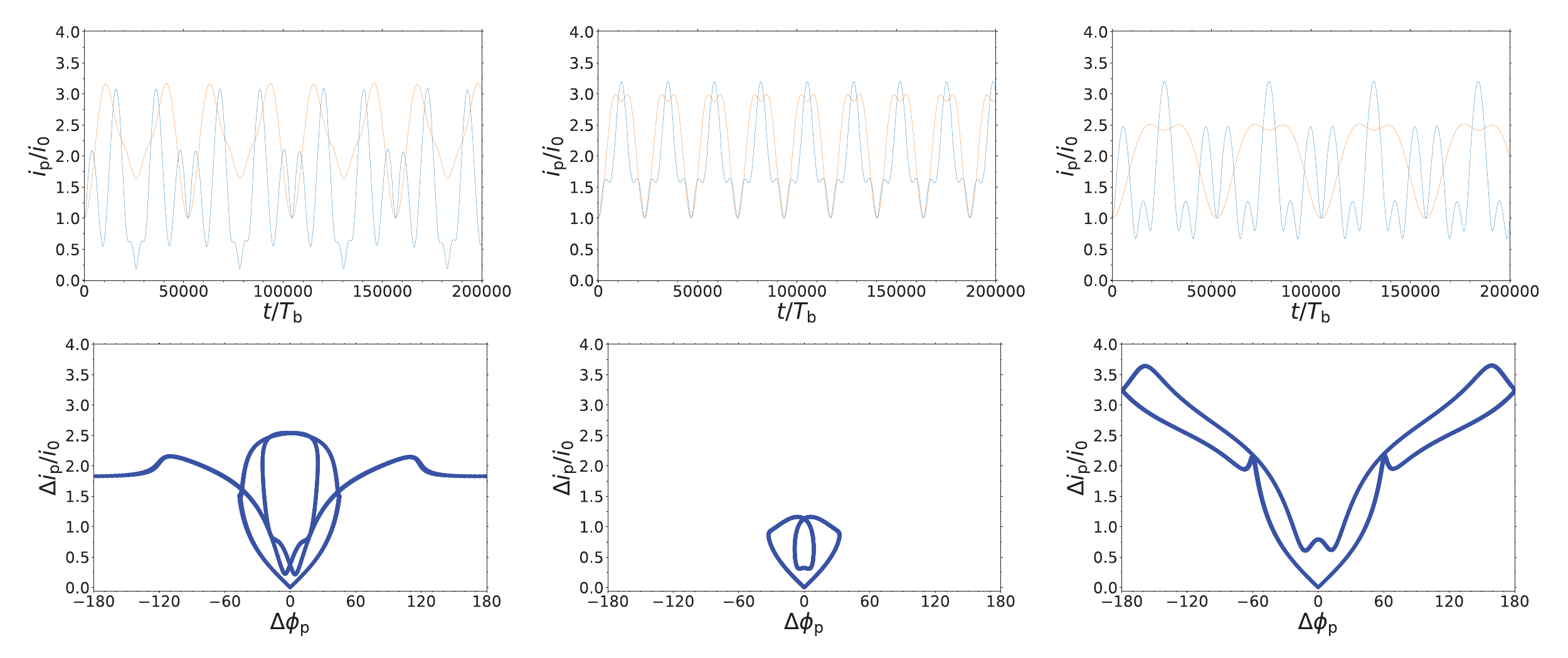}
\caption{Analytic models for planets that are in secular resonance: T1 (left), T2 (middle) and T3 (lower). Time evolution of planet inclinations (upper panels) and the relative tilt between two planets as a function of their nodal phase angle difference (lower panels). }
\label{aT1}
\end{figure*}

Figure~\ref{aT1} shows some examples of systems that are in resonance based on the analytic model. In the left panels we show an example where the planets display both circulation and libration in the resonance. The two planets are at semi-major axes of 11.4 and 17.0$\,a_{\rm b}$ with $i_{\rm p}$ = 1$^\circ$ and $e_{\rm b}$ = 0.8. With these parameters, the ratio $\omega_1$/$\omega_2$ = 4 in the analytic solution. In the upper panel, the inclinations of the inner and outer planets display complex but exactly periodic oscillations. In the lower panel, the phase diagram shows that this system undergoes both libration and  circulation. Moreover, unlike other plots that show both libration and circulation,  the plot for the resonance case displays a complex but orderly pattern.

In the middle panels of Figure~\ref{aT1}, we show an example of a system that is in resonance and librating. The two planets are at semi-major axes of 12.0 and 16.02 $a_{\rm b}$
with $i_0$ = 1$^\circ$ and  the same binary eccentricity $e_{\rm b} = 0.8$.  For this case, the ratio $\omega_1$/$\omega_2$ = 5 in the analytic solution. The two planets are locked to each other in nodal phase angle during the tilt oscillations, while $i_{\rm p}/i_0$ for both planets are always above 1. In the lower panel,  unlike other plots of libration which we showed above have an inverted triangle. The libration resonance displays a big oscillation and a small oscillation in each period.

In the right panels of Figure~\ref{aT1} we show an example of a system that is in resonance and circulating, model T3. The system has the same ratio $\omega_1$/$\omega_2$ = 5 as model T2 shown in the middle panels. The model parameters are the same as model T2, except that the semi-major axis of the outer planet is $a_{\rm p2}$ = 21.18$\,a_{\rm b}$.   The lower-right panel is unlike any of the 
plots of circulation about a circular orbit binary that involve simple V-shapes.

\begin{figure}
\includegraphics[width=8.7cm]{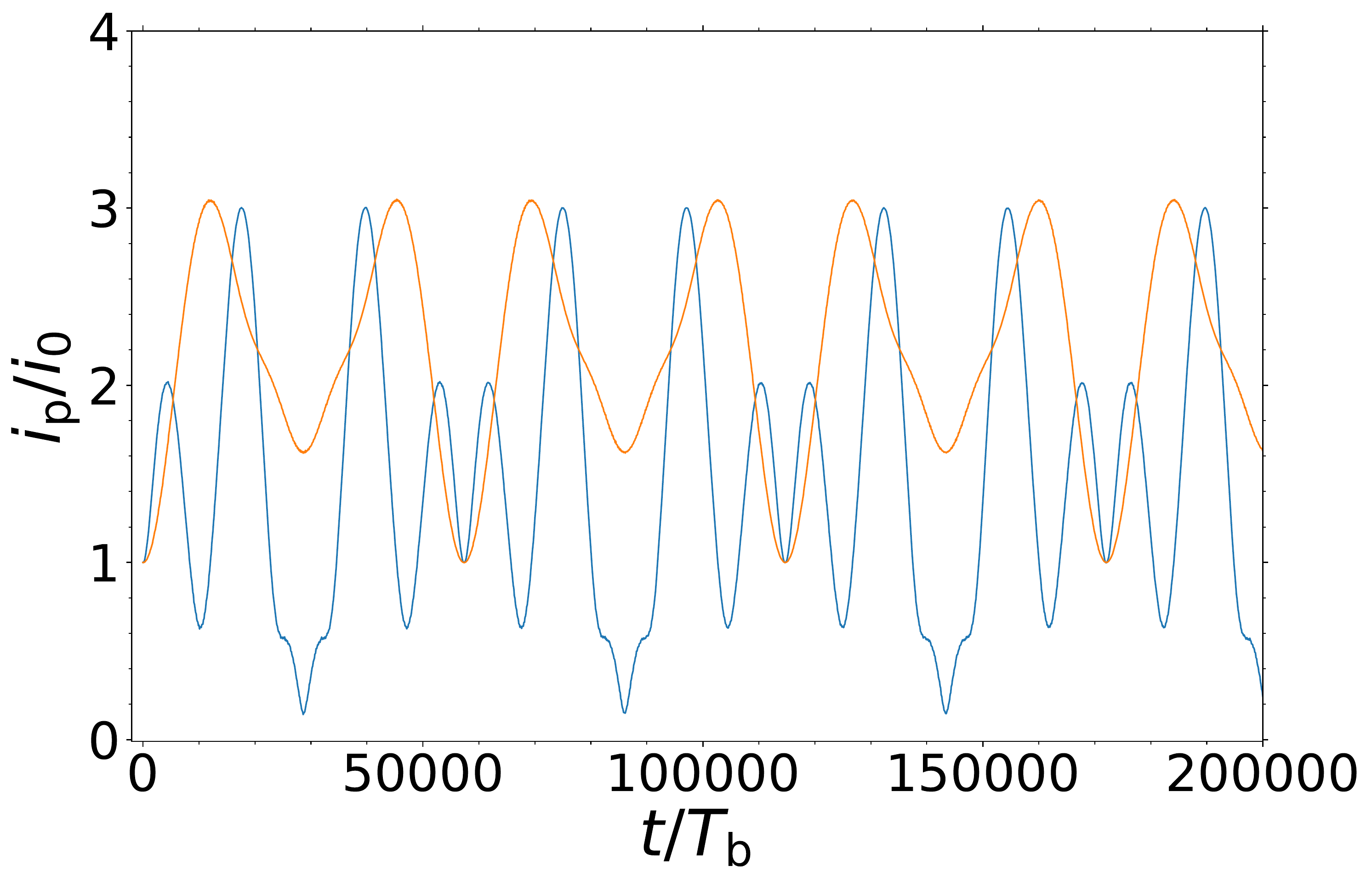}
\includegraphics[width=8.7cm]{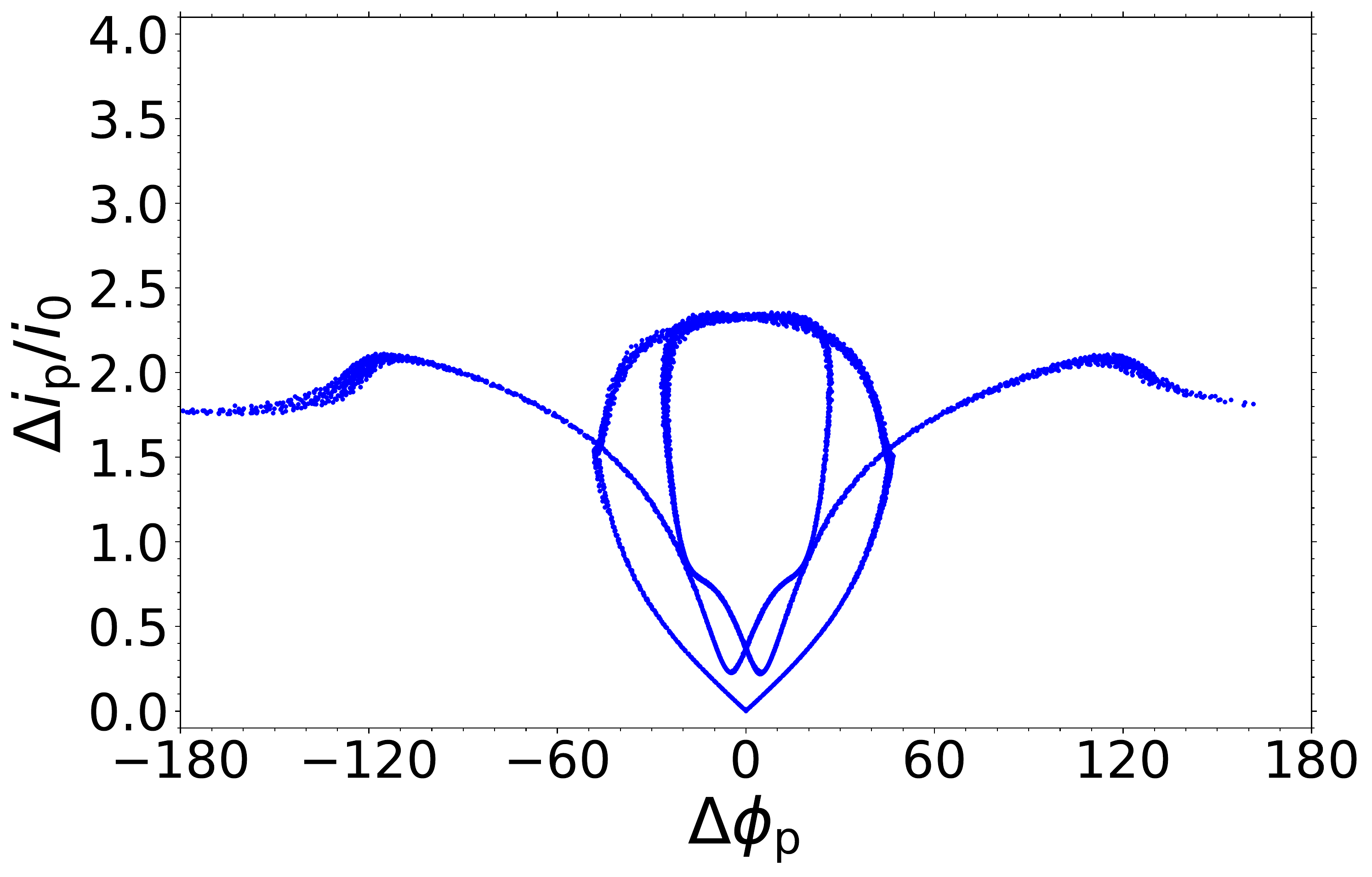}
\caption{Numerical model of planets that are in resonance: same as model T1 except $a_{\rm p1}$ = 12$a_{\rm b}$ and $a_{\rm p2}$ = 18$a_{\rm b}$.} 
\label{nT1}
\end{figure}

Numerical simulations also show resonances, but for somewhat different parameters 
than predicted by the analytic model. 
In Figure~\ref{nT1}, we plot a resonance case based on numerical simulations
that is nearly identical to the model T1 in Figure~\ref{aT1}, but involves somewhat different planet
orbital radii of $a_{\rm p1}$ = 12$a_{\rm b}$ and $a_{\rm p2}$ = 18$a_{\rm b}$. It displays the same resonance behaviour as the analytic results, although the period is longer than the analytic results.  The phase plots of $\Delta i/i_0$
as a function of $\Delta \phi_{\rm p}$ are nearly identical.
We attempted to reproduce the resonances for models T2 and T3 using numerical
simulations but we were
unable to find these cases, since we do not have an efficient algorithm
for finding them. However, we believe they exist.

\section{Secular Evolution of nearly polar circumbinary planets}
\label{high}

In this section we consider the evolution of a circumbinary planetary system in which two planets are initially mutually coplanar, but on nearly polar orbits with respect to the binary. 
Eccentric binaries are more likely to host circmbinary discs that are highly misaligned with respect to the binary \citep[e.g.,][]{Czekala2019}. We expect that
should be also true for circumbinary planets.
We investigate the evolution of two circumbinary planets in a binary system with $e_{\rm b}$ = 0.5 using an analytic secular model and numerical simulations.
For a single planet orbiting a binary, there is a stationary inclination for which
 the binary and the planet precess together with a constant relative tilt \citep{Farago2010,MartinandLubow2019}.  The stationary planet
 tilt angle with respect to the binary orbital plane is $90^\circ$ if the planet angular momentum is very small compared to the binary
 angular momentum. Otherwise the stationary tilt is reduced. This situation is unlike the nearly coplanar case, where the stationary tilt is always zero, independent of the planet angular momentum.
 Consequently, there is a basic difference in the behaviour of the orbital evolution of a planet
 in a nearly polar state versus in a nearly coplanar state. 
 
 An analytic treatment of the evolution of two nearly polar circumbinary planets requires
 consideration of their stationary configuration, as well as oscillations away from that configuration.
In a linear model for the nearly polar planets case, we must account for the change
in the stationary tilt angle away from 
$90^\circ$. When the initial planet tilts differ from
the stationary tilts, the planet orbits undergo tilt oscillations.

The equations are analysed
in a frame that precesses with the binary such that the $x$-direction remains along the
instantaneous direction of the binary eccentricity and the $z$-direction remains along the binary
angular momentum. The binary precession is due to its gravitational interaction
with the two planets.
This precessing frame has the Cartesian axes ($\bm{e}_{\rm b}$, $\bm{\ell}_{\rm b}\times \bm{e}_{\rm b}$, $\bm{\ell}_{\rm b}$), where $\bm{\ell}_{\rm b}$ is binary tilt vector that is the unit
 vector that is parallel to the binary angular momentum, $\bm{J}_{\rm b}$. 
In this precessing frame, we have that $\bm{\ell}_{\rm b}=(0,0,1)$ at all times. 

 Denoting the planet tilt vector for planet $j$ as $\bm{\ell}_{\rm pj}$, we calculate the inclination of its orbital plane relative to the orbital plane of the binary with
\begin{equation}
i_{\rm pj} =\cos^{-1}(\bm{\ell}_{\rm b}\cdot \bm{\ell}_{\rm pj}).
\label{ipp}
\end{equation}
The longitude of ascending node of the planet $j$ in a frame relative to the instantaneous  angular momentum and eccentricity vectors of the binary is
\begin{equation}
\Phi_{\rm pj}=\tan^{-1}\left(\frac{ -\bm{\ell}_{\rm pj}\cdot \bm{e}_{\rm b}} { \bm{\ell}_{\rm pj}\cdot (\bm{\ell}_{\rm b}\times \bm{e}_{\rm b})} \right).
\label{Phipp}
\end{equation}
Equations (\ref{ipp}) and (\ref{Phipp}) reduce to the same 
the corresponding expressions as Equations (\ref{ipn})
and (\ref{eq:phi}) respectively in the nearly coplanar case if the precession
of the binary is ignored so that $\bm{\ell}_{\rm b}$ is along the $z-$direction in the inertial frame.

We determine equations for the planet tilts $(\ell_{\rm p1x}, \ell_{\rm p1y}, \ell_{\rm p1z})$ in the frame that is precessing 
with the binary.
We assume that
$\ell_{\rm p1x}  \simeq \ell_{\rm p2x} \simeq \pm 1$ for both planets and that 
$|\ell_{\rm p1y}|,|\ell_{\rm p1z}|, |\ell_{\rm p2y}|,|\ell_{\rm p2z}| \ll 1$. The tilt evolution equations that are linearized in $\ell_{\rm p1y},\ell_{\rm p1z}, \ell_{\rm p2y}, \ell_{\rm p2z}$ are
\begin{gather}
s \, J_{\rm p1} \frac{d \ell_{\rm p1y}}{dt} =  C_{\rm p1, p2} ( \ell_{\rm p1z} -  \ell_{\rm p2z}) + \tau_{\rm y} \omega_{\rm p1} \ell_{\rm p1z} J_{\rm p1} +   3 \sqrt{1-e_{\rm b}^2} \, ( \omega_{\rm p1} \gamma_{\rm p1} + 
\omega_{\rm p2} \gamma_{\rm p2})   J_{\rm p1},
\label{eqpolary1} \\
s \, J_{\rm p1} \frac{d \ell_{\rm p1z}}{dt} =  C_{\rm p1, p2} ( \ell_{\rm p2y} -  \ell_{\rm p1y}) + \tau_{\rm z} \omega_{\rm p1} \ell_{\rm p1y} J_{\rm p1},
\label{eqpolaryz1} \\
s \, J_{\rm p2} \frac{d \ell_{\rm p2y}}{dt} =  C_{\rm p1, p2} ( \ell_{\rm p2z} -  \ell_{\rm p1z}) + \tau_{\rm y} \omega_{\rm p2} \ell_{\rm p2z} J_{\rm p2}
   +   3  \sqrt{1-e_{\rm b}^2} \, ( \omega_{\rm p1} \gamma_{\rm p1} + 
 \omega_{\rm p1} \gamma_{\rm p2}) J_{\rm p2},
\label{eqpolary2} \\
s \, J_{\rm p2} \frac{d \ell_{\rm p1z}}{dt} =  C_{\rm p1, p2} ( \ell_{\rm p1y} -  \ell_{\rm p2y}) + \tau_{\rm z} \omega_{\rm p2}\ell_{\rm p2y} J_{\rm p2},
\label{eqpolarz2} \\
s\, \frac{d e_{\rm b}}{dt} = 5  e_{\rm b} \sqrt{1- e_{\rm b}^2} \, ( \omega_{\rm p1} \gamma_{\rm p1} \ell_{\rm p1y} + 
 \omega_{\rm p1} \gamma_{\rm p2} \ell_{\rm p2y},
 ) 
\label{eqpolare}
\end{gather} 
where  $s= \sgn{(\ell_{\rm p1x})}=\sgn{(\ell_{\rm p2x})}$, $C_{\rm p1, p2}$ is the planet-planet interaction coefficient that is defined by Equation (\ref{eq:pp}), 
$\omega_{\rm p1}$ and $\omega_{\rm p2}$ are defined in Equations (\ref{omegap1}) and (\ref{omegap2}),
respectively, and the binary torque coefficient $\tau_{\rm y}$ is given
by Equation (\ref{tauyp}) and
\begin{equation}
    \tau_{\rm z} =  5e_{\rm b}^2, \label{tauzp}
\end{equation}
and 
\begin{gather}
    \gamma_{\rm p1} = \frac{\sqrt{1-e_{\rm b}^2} J_{\rm p1}}{J_{\rm b}}, \label{gammap1} \\
    \gamma_{\rm p2} = \frac{\sqrt{1-e_{\rm b}^2} J_{\rm p2}}{J_{\rm b}}, \label{gammap2} 
\end{gather}
where $J_{\rm b}$ is the angular momentum of the binary.
The final terms on the RHSs of Equations (\ref{eqpolary1}) and 
(\ref{eqpolary2})
are due to the precession of the orbit of the 
binary caused by its interaction with the planets.

We now assume that $\gamma_{\rm p1}, \gamma_{\rm p2} \ll 1 $ and are of the same order
as the $y$ and $z$ tilt component magnitudes $|\ell_{\rm p1y}|,|\ell_{\rm p1z}|, |\ell_{\rm p2y}|,|\ell_{\rm p2z}|$. To linear order, we then have that
from Equation (\ref{eqpolare}) that
\begin{equation}
\frac{d e_{\rm b}}{dt} =0 \label{debdt0}
\end{equation}
and therefore regard $e_{\rm b}$ as a constant and solve Equations (\ref{eqpolary1}) -(\ref{eqpolarz2}).

Notice that Equations (\ref{eqpolary1}) -  (\ref{eqpolarz2}) contain terms that are 
linear in the tilt components, as well as terms that are independent of the tilt components.
To satisfy these equations, we write the solution in the form
\begin{equation}
\ell_{\rm p1y}(t) = \ell_{\rm p1ys} +  \ell_{\rm p1yv}(t), \label{ellform}
\end{equation}
where $\ell_{\rm p1ys}$ is independent of time and represents the stationary part of the solution
and $\ell_{\rm p1yv}(t) ={\rm Re}[\ell_{\rm p1yv}\exp(i\omega t)]$ represents the time-dependent part of the solution.
The other components of the tilt vectors are decomposed in a similar way.

\subsection{Stationary inclinations of two nearly polar circumbinary planets}

We determine the stationary solutions to  Equations (\ref{eqpolary1}) - (\ref{eqpolarz2})  by
using a solution of the form of Equation (\ref{ellform}). 
The LHSs of Equations (\ref{eqpolary1}) - (\ref{eqpolarz2}) are zero.
From Equations (\ref{eqpolaryz1}) and (\ref{eqpolarz2}), it then follows that the $y$-components of the stationary tilt vectors are zero. That is,
\begin{equation}
\ell_{\rm p1ys} = \ell_{\rm p2ys} = 0. \label{ellp1ys}
\end{equation}
From Equations (\ref{eqpolary1}) and (\ref{eqpolary2}), it then follows that the $z$-components of the stationary tilt vectors are given by 
\begin{equation}
\ell_{\rm p1zs} = \frac{
3\sqrt{1-e_{\rm b}^2} \left[
(1+4e_{\rm b}^2) 
\omega_{\rm p2} J_{\rm p1} J_{\rm p2} - C_{\rm p1, p2} (J_{\rm p1} + J_{\rm p2}) \right]
( \omega_{\rm p1} \gamma_{\rm p1}+ \omega_{\rm p2} \gamma_{\rm p2})
}
{
(1+4e_{\rm b}^2) \left[ (1+4e_{\rm b}^2) \omega_{\rm p1} \omega_{\rm p2} J_{\rm p1}J_{\rm p2} - C_{\rm p1, p2} (\omega_{\rm p1} J_{\rm p1} + \omega_{\rm p2} J_{\rm p2} ) \right] \label{ellp1zs}
} 
\end{equation}
and
\begin{equation}
\ell_{\rm p2zs} = \frac{
3\sqrt{1-e_{\rm b}^2} \left[
(1+4e_{\rm b}^2) 
\omega_{\rm p1} J_{\rm p1} J_{\rm p2} -  C_{\rm p1, p2} (J_{\rm p1} + J_{\rm p2}) \right]
( \omega_{\rm p1} \gamma_{\rm p1}+ \omega_{\rm p2} \gamma_{\rm p2})
}
{
(1+4e_{\rm b}^2) \left[ (1+4e_{\rm b}^2) \omega_{\rm p1} \omega_{\rm p2} J_{\rm p1}J_{\rm p2} - C_{\rm p1, p2} (\omega_{\rm p1} J_{\rm p1} + \omega_{\rm p2} J_{\rm p2} ) \right]
}.  \label{ellp2zs}
\end{equation}

Combining Equations (\ref{ellp1ys}) - (\ref{ellp2zs}), we have that the stationary orbit inclinations
are given by
\begin{gather}
i_{\rm p1s} = \arccos{(\ell_{\rm p1zs})} \label{eqi} \\
i_{\rm p2s} = \arccos{(\ell_{\rm p2zs})} \label{eqo}, 
\end{gather}
with $\ell_{\rm p1zs}$ and $\ell_{\rm p2zs}$ given by Equations (\ref{ellp1zs}) and (\ref{ellp2zs}),
respectively.

Equations (\ref{ellp1zs}) and (\ref{ellp2zs}) for the stationary tilt components $\ell_{\rm p1zs}$ and  $\ell_{\rm p2zs}$ contain the factor
$( \omega_{\rm p1} \gamma_{\rm p1}+ \omega_{\rm p2} \gamma_{\rm p2})$. This factor appears in the terms of  Equations (\ref{eqpolary1}) and (\ref{eqpolary2}) that occur because of the precession of
binary due to the gravitational effects of the planets. This factor is zero if the planets
are massless in which case $\ell_{\rm p1zs}= \ell_{\rm p2zs}=0$. From Equations (\ref{eqi}) and 
(\ref{eqo}), it follows that the stationary planet inclinations are perpendicular to the orbital plane of the binary, as occurs for test particles \citep{Farago2010}. 
In the case that $J_{\rm b} \gg J_{\rm p1}>0$ and $J_{\rm p2} =0$,
Equations (\ref{eqi}) and (\ref{eqo}) reduce to equation 
19 of \cite{MartinandLubow2019} for the stationary tilt of a low angular momentum single planet (planet 1) that orbits a binary.

\begin{figure}
\includegraphics[width=8.6cm]{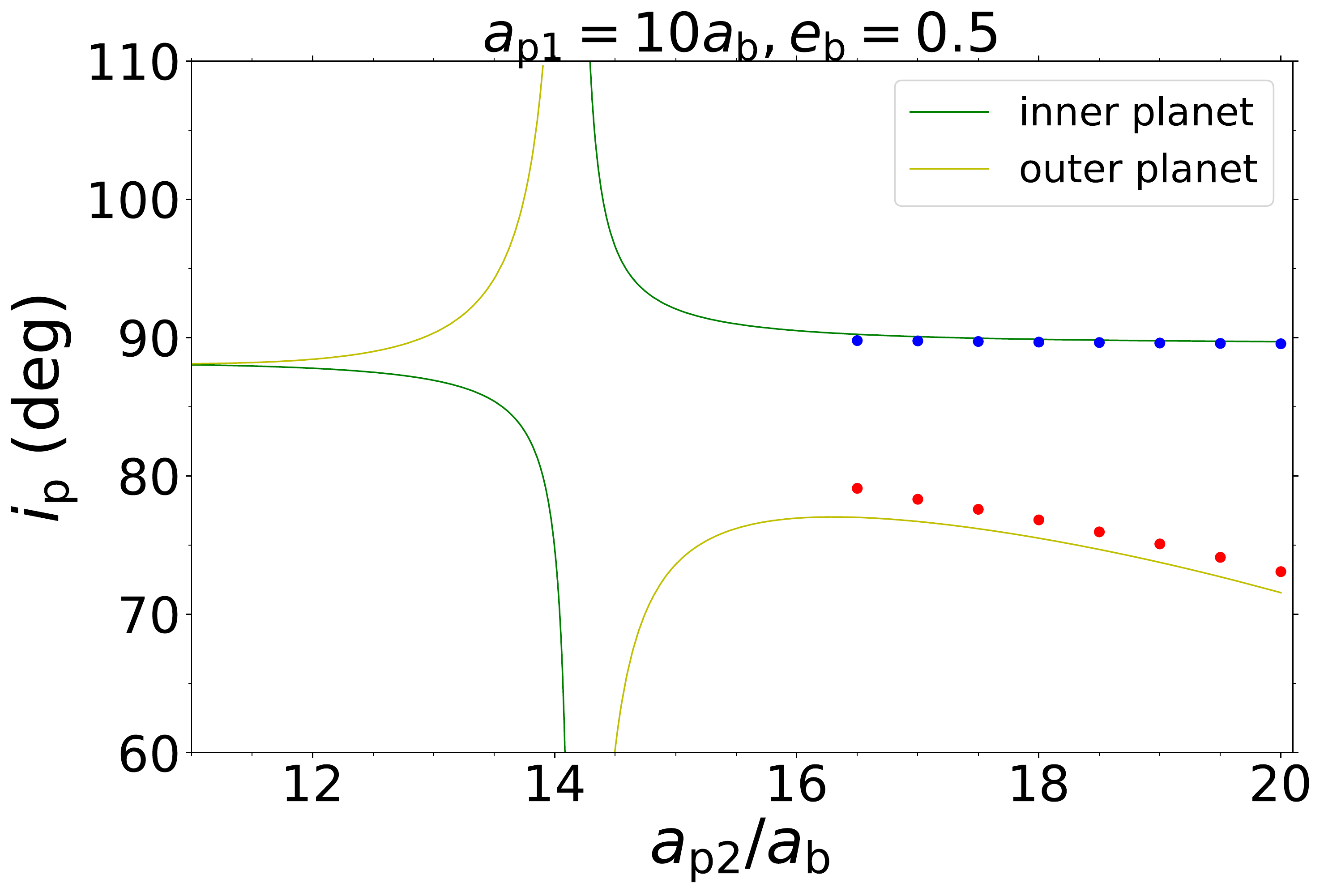}
\caption{Stationary inclinations of the inner and outer planet in the binary system with  $e_{\rm b}$ = 0.5. We fix $a_{\rm p1} = 10 a_{\rm b}$
and vary $a_{\rm p2}$ from 11 to 20 $a_{\rm b}$. Green and yellow lines are the analytic solutions  of the inner and outer planet given respectively given by Equations (\ref{eqi}) 
and (\ref{eqo}). The blue dots and red dots are numerical determinations of the stationary inclinations by means of simulation results.} 

\label{istat}
\end{figure}

In Figure~\ref{istat}, we plot the analytic stationary solutions given by Equations (\ref{eqi}) 
and (\ref{eqo}) 
for the inner and the outer planets (green and yellow lines) in a binary system with $e_{\rm b}$ = 0.5. We fixed the inner planet semi-major axis to 10 $a_{\rm b}$ while the semi-major axis of the outer planet ranges from $11 a_{\rm b}$ to  $20 a_{\rm b}$. There is a resonance at $a_{\rm p2}=14.2 a_{\rm b}$ where
the denominators vanish in Equations (\ref{ellp1zs}) 
and (\ref{ellp2zs}). Near this value,
we see in the plot that the stationary tilt changes rapidly.

In addition, we  determine the stationary
inclinations by means of four body simulations (the blue and red dots) to compare with analytic solutions. To determine the stationary
inclinations in the four body simulations, we first run a simulation with the initial values of $i_{\rm p1s}$ and $i_{\rm p2s}$ calculated from the analytic models. Since $\bm{\ell}_{\rm p1}$ and $\bm{\ell}_{\rm p2}$ precess about their stationary inclinations in the simulation, we can find the maximum and minimum inclination $i_{\rm p}$ at $\Phi_{\rm p}$ = 0 (see Figure~\ref{iphase}). We then determine the mean value of the maximum and minimum of $i_{\rm p}$ for each planet and use these values as the initial conditions for the next simulation. We iterate on the initial conditions in this way four times and after this we find that both planets are very close to stationary.  The results for the inner planet in blue dots are consistent with the green line. The results for the outer planet in red dots lie slightly above the yellow  line.  


 We consider a sequence of models in which the planet mass
 for both planets changes by the same scale factor $S$.
 To lowest order in $0 \le S \ll 1$, the difference in stationary tilts 
 is  given by
 \begin{equation}
     \ell_{\rm p2zs} - \ell_{\rm p1zs} = 
     \frac{3\sqrt{1-e_{\rm b}^2}(\omega_{\rm p1}-\omega_{\rm p2})
      (
      \omega_{\rm p1} \gamma_{\rm p1}+ \omega_{\rm p2} \gamma_{\rm p2}
      ) }{ (1+4e_{\rm b}^2) \omega_{\rm p1} \omega_{\rm p2} }\, S.
 \end{equation}
 For $S=0$ both planet orbits are aligned, as expected since they
 both have stationary tilt angles of $90^\circ$. But as the planet
 masses increase, they become misaligned in a stationary configuration
 due to the planet interactions with the binary, even though
 their mutual gravitational interaction increases.

\subsection{Tilt oscillations of two nearly polar circumbinary planets}

We determine the equations for the time-dependent contributions to Equation (\ref{ellform}) using Equations (\ref{eqpolary1}) - (\ref{eqpolarz2}) in a similar manner as the 
coplanar case with Equations (\ref{i1}) - (\ref{i4}) and obtain
\begin{gather}
 i \omega \,  s\, J_{\rm p1}  \ell_{\rm p1yv}  =  C_{\rm p1, p2} ( \ell_{\rm p1zv} -  \ell_{\rm p2zv}) + \tau_{\rm y} \omega_{\rm p1} \ell_{\rm p1zv} J_{\rm p1},
\label{eqpolary1v} \\
i \omega \,  s\, J_{\rm p1}  \ell_{\rm p1zv} =  C_{\rm p1, p2} ( \ell_{\rm p2yv} -  \ell_{\rm p1yv}) + \tau_{\rm z} \omega_{\rm p1} \ell_{\rm p1yv} J_{\rm p1},
\label{eqpolaryz1v} \\
i \omega \, s\, J_{\rm p2}  \ell_{\rm p2yv}  =  C_{\rm p1, p2} ( \ell_{\rm p2zv} -  \ell_{\rm p1zv}) + \tau_{\rm y} \omega_{\rm p2} \ell_{\rm p2zv} J_{\rm p2},
\label{eqpolary2v} \\
i \omega \,  s\, J_{\rm p2}  \ell_{\rm p1zv} =  C_{\rm p1, p2} ( \ell_{\rm p1yv} -  \ell_{\rm p2yv}) + \tau_{\rm z} \omega_{\rm p2}\ell_{\rm p2yv} J_{\rm p2},
\label{eqpolarz2v} 
\end{gather}
where again $s= \sgn{(\ell_{\rm p1x})}=\sgn{(\ell_{\rm p2x})}$ and $\omega$ is the oscillation frequency and $\tau_y$ and $\tau_z$ are given by Equations (\ref{tauyp}) and (\ref{tauzp}), respectively. As in the coplanar case, we solve these equations using normal modes subject to the initial conditions.

\begin{figure*}
\includegraphics[width=8.6cm]{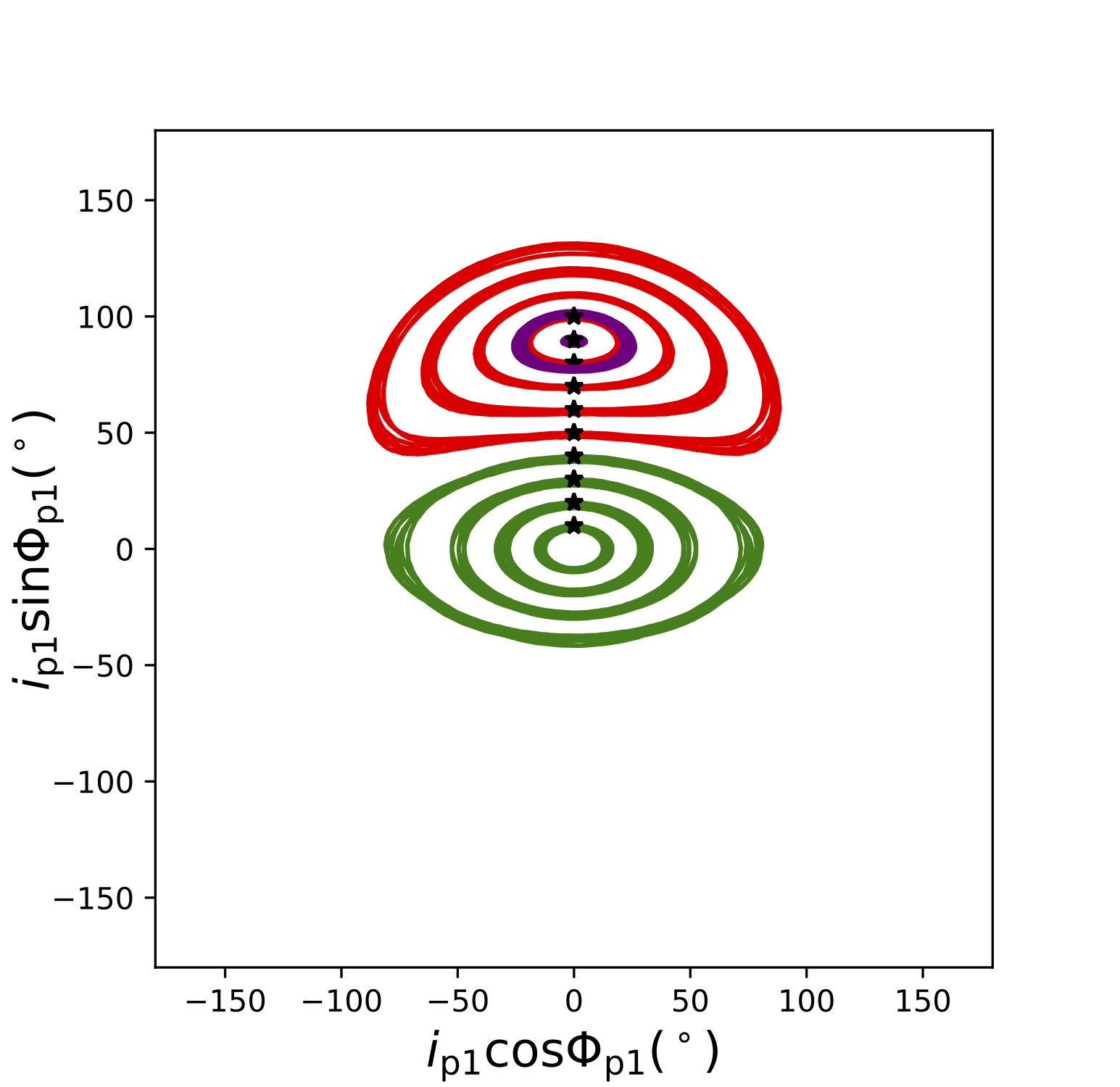}
\includegraphics[width=8.6cm]{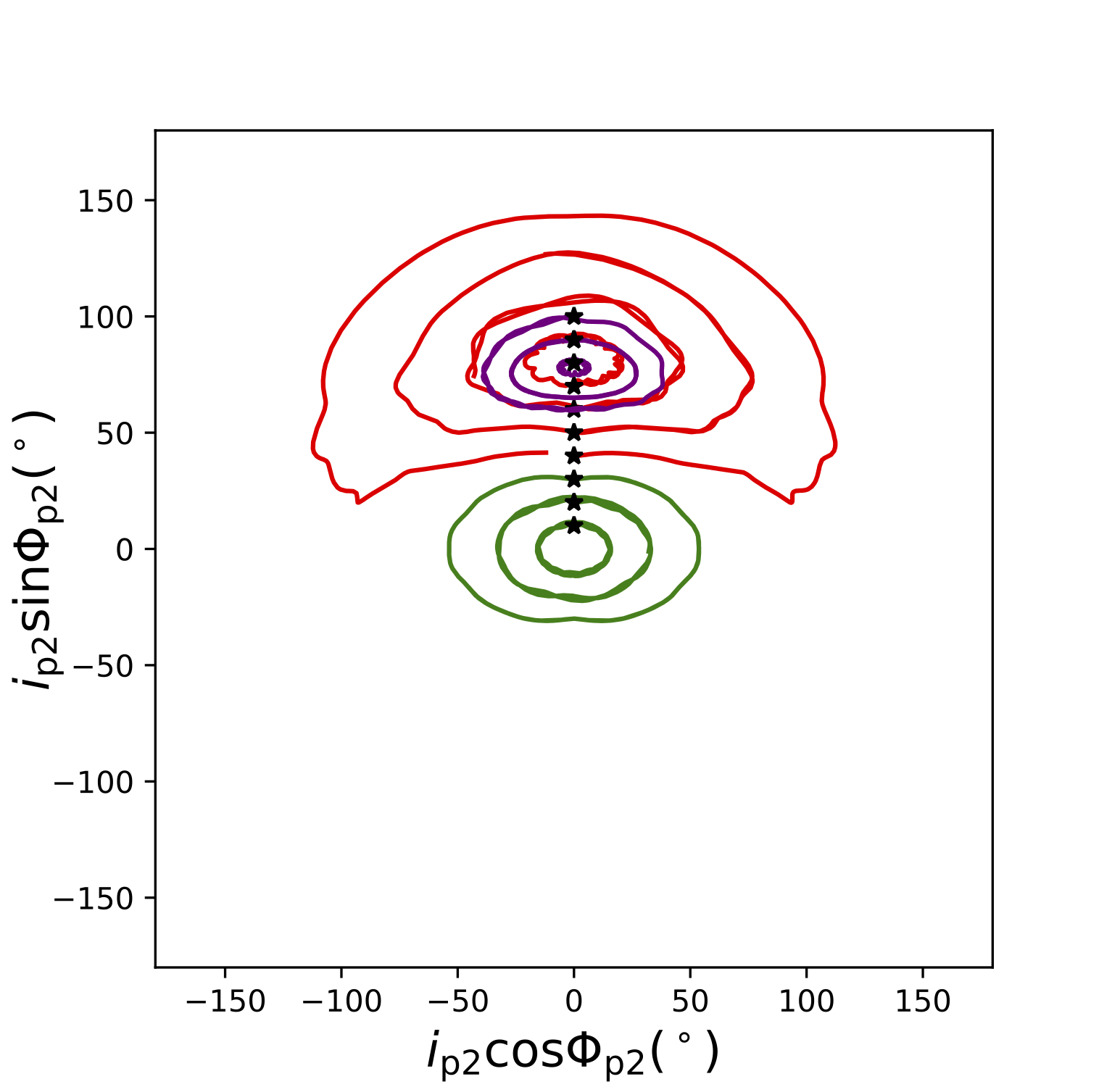}

\caption{The $i_{\rm p}\cos \Phi_{\rm p}-i_{\rm p} \sin \Phi_{\rm p}$ phase plane of models similar to H1 for orbits with various  initial inclinations ranging from 10$^\circ$ to 100$^\circ$ with  10$^\circ$ interval. The left (right) panel is for the inner (outer) planet. Green lines are prograde circulating orbits, red lines are polar librating orbits with initial inclination $i_{\rm p}$ < $i_{\rm s}$ and purple lines are polar librating orbits with initial inclination $i_{\rm p}$ > $i_{\rm s}$. The initial positions are shown by the black stars. } 
\label{iphase}
\end{figure*}

To understand the tilt oscillations between two planets that are highly inclined
with respect to the binary, we consider in model H1  two planets with initial  inclination $80^\circ$, semi-major axes 10 $a_{\rm b}$ and 18 $a_{\rm b}$ around a binary with eccentricity $e_{\rm b}$ = 0.5.
According to the previous subsection, we know that the stationary inclination of the inner planet is $89.7^{\circ}$ and the stationary inclination of the outer planet is $76.8^{\circ}$. Figure~\ref{iphase} shows the $i_{\rm p} \sin \Phi_{\rm p}-i_{\rm p}\cos \Phi_{\rm p}$ phase plane of the inner planet (left panel) and the outer planet (right panel) that have the same properties as model H1 but with initial inclination ranging from 10$^\circ$ to 100$^\circ$.  The different colours represent different types of orbits. The green lines represent prograde circulating orbits, the red lines represent polar librating orbits with initial inclination  $i_{\rm p}$ < $i_{\rm s}$ and the purple lines represent polar librating orbits with initial inclination $i_{\rm p}$ > $i_{\rm s}$. The stationary inclination for the inner planet,  $i_{\rm s}$, is close to 90$^\circ$ in the left panel, while  the stationary inclination for the outer planet is between 70$^\circ$ and 80$^\circ$ in the right panel. These plots are consistent with the stationary inclinations shown in Figure~\ref{istat}. The fact that the lines are bold and somewhat irregular is likely due to the effects of planet-planet interactions.

\begin{figure*}
\includegraphics[width=8.6cm]{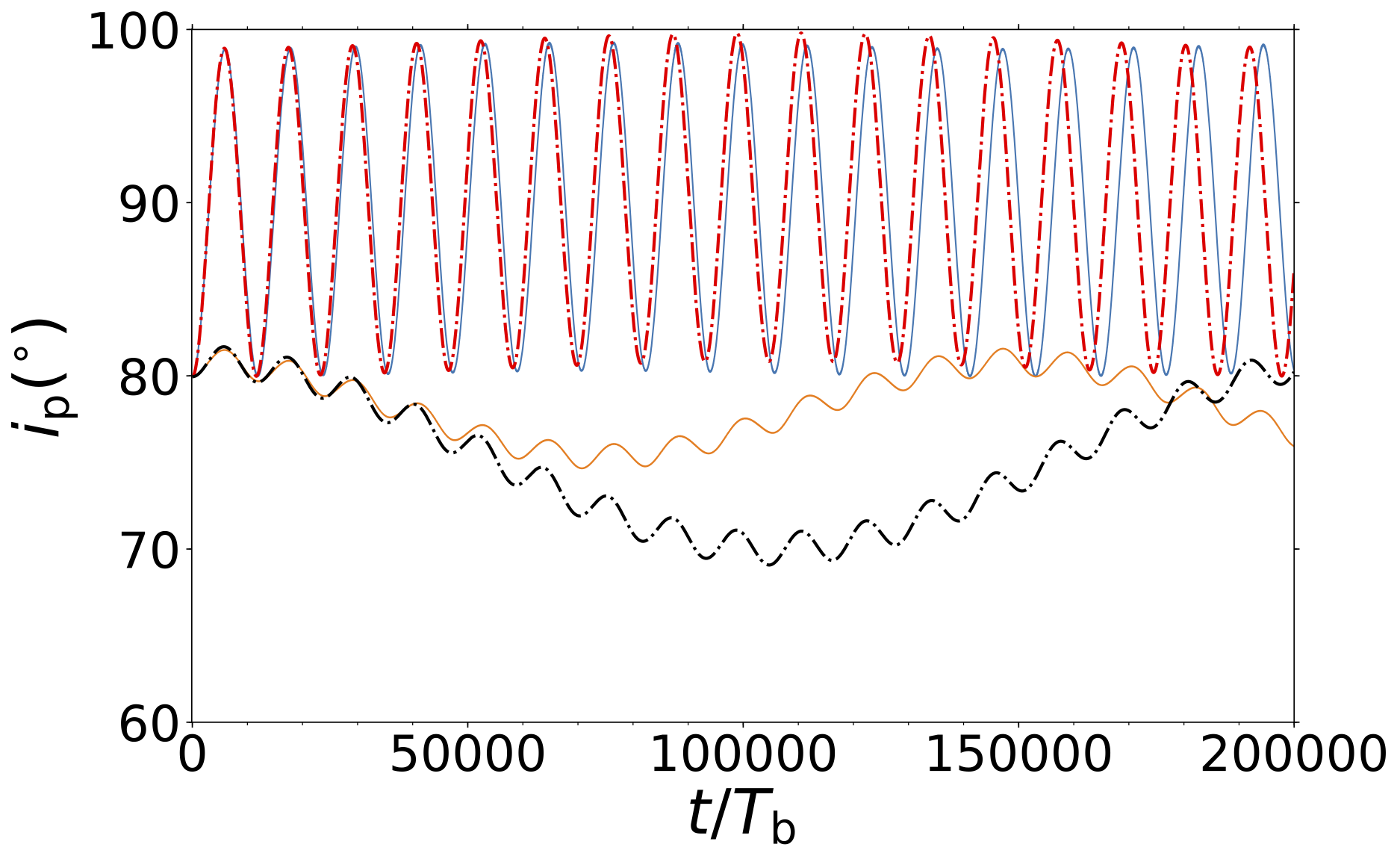}
\includegraphics[width=8.6cm]{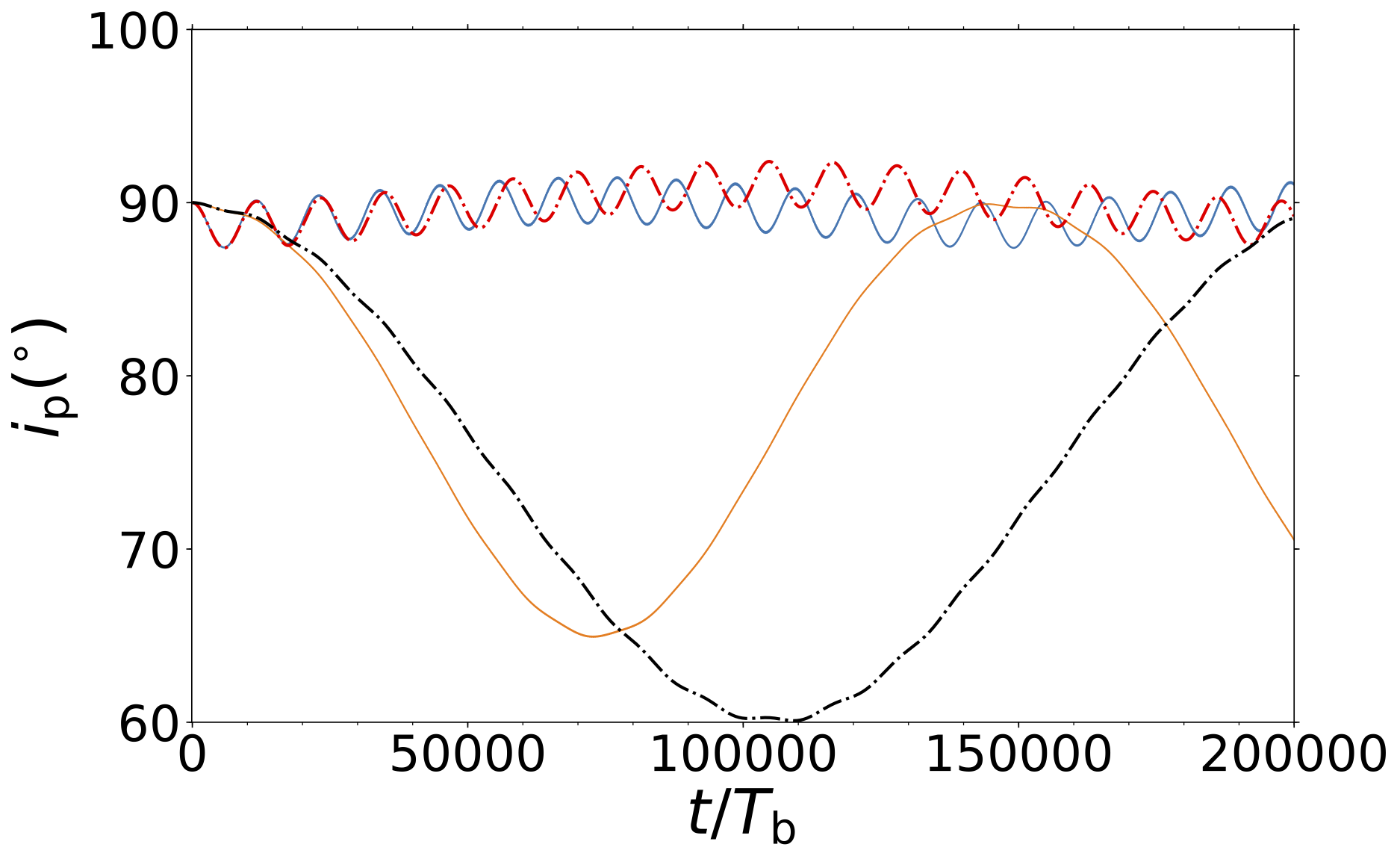}

\caption{Time evolution of the planet inclinations $i_{\rm p}$ for Model H1 in which the planets have initial inclination of $80^\circ$ (left panel) and H2 in which the planets have an initial inclination of $90^\circ$ (right panel). The solid lines represent the numerical simulation results of the inner planet (blue) and the outer planet (yellow). The dash-dotted lines represent the analytical results of the inner planet (red) and the outer planet (black). }
\label{H1H2}
\end{figure*}

The left panel of Figure~\ref{H1H2} shows the time evolution of the planet inclinations $i_{\rm p1}$ and $i_{\rm p2}$ for model H1 in which the planets begin at an inclination of $80^\circ$. The blue and yellow solid lines correspond to the inner and outer planets, respectively, and the red and black dash-dotted lines are the analytic solutions of the inner and the outer planets, respectively.  The polar libration period of the inner planet (blue solid line) is about 10000 $T_{\rm b}$ and its inclination $i_{\rm p1}$ oscillates from 80$^\circ$ to 98$^\circ$. 
On the other hand, the polar libration period of the outer planet is about 150000 $T_{\rm b}$ and its inclination $i_{\rm p2}$ decreases from 80$^\circ$ to 73$^\circ$. Furthermore, like Models A1, A2 and A3, the two lines display two behaviours in the time evolution. The small peaks with a period of about 10000 $T_{\rm b}$ shown in the solid yellow line  indicate the tilt oscillations due to the polar libration of the inner planet, while the long term behaviour in the  blue line is slightly affected by the tilt oscillation of the outer planet.  

The right panel of Figure~\ref{H1H2} shows the time evolution of the planet inclinations for model H2 with the two planets initially at an inclination of 90$^\circ$. Because they begin at an inclination that is very close to the stationary inclination  of the inner planet, the amplitude of the polar libration is much smaller than that in model H1. Thus, its effect on the oscillation of the outer planet is not evident in the solid yellow line. However, the outer planet is much farther from its stationary inclination and so the amplitude of the polar libration is much larger than that in model H1. Nevertheless, the long term behaviour of the inner planet (blue line) shows a small effect at the period of the outer planet.    

The analytical and numerical results agree well for the inner planet in model H1
(red dashed-dotted line and blue solid line) over the entire time range considered. There is good agreement for the inner planet in model
H2 until a time of about 70000 $T_{\rm b}$. Beyond that time the oscillation periods differ and the oscillation centres differ, while the oscillation amplitudes are similar.
There are large deviations between the analytical results and the numerical results for
the outer planet over long timescales.
The analytic and numerical results for the outer planet in model H1 (black dashed-dotted line and yellow solid line) agree well until a time of about 50000 $T_{\rm b}$. We have investigated the possible origin of this discrepancy.
The analytic model assumes that the relative tilt between the planets
is small in calculating the planet-planet interaction term and that the planet
inclination is close to a polar configuration in the planet-binary interaction term.
These assumptions likely break down after some evolution of these
planets from their initially mutually coplanar and nearly polar states.
This effect is stronger on the outer planet because it is more affected
by the companion planet than the binary and deviates more from the polar configuration. The inner planet on the other hand is more
dominated by the effects of the binary. It also remains closer to being polar with the binary.
 In the right panel of Figure~\ref{H1H2}, the deviation between the analytic and the numerical results can be seen after about 20000 $T_{\rm b}$. Again, the discrepancy may be due to a breakdown of the assumptions in the analytic model.

\section{Discussion and Conclusions} 
\label{dis}

We have investigated the orbital dynamics of circumbinary planetary systems with two planets that are on inclined orbits 
around a circular or eccentric orbit binary. 
We considered planet orbits that are initially circular and coplanar to each other, but misaligned with respect to the binary orbital plane.
We examined cases of small initial planet orbit misalignments with respect to the binary orbital plane,
as well as large initial planet orbit misalignments (almost polar) with respect to the binary orbital plane.
The joint effects of planet-planet interactions and binary-planet  interactions can result in complex planet tilt oscillations.
We used analytic models and numerical simulations to explore the effects changing the values of the 
planet semi-major axes, binary eccentricity, and initial inclination.

 In the case that the planet orbital planes are nearly aligned with the orbital plane
of a circular orbit binary, the secular tilt oscillations are driven only by planet-planet interactions.
The circular orbit binary does not drive secular tilt oscillations.
 The tilt oscillation frequency is the same for both planets and the tilt oscillations are periodic (see Figures~\ref{aA0} to \ref{aC2}).
In  such cases, the two planets undergo mutual libration  if they are close together and circulation they are if far
apart with an abrupt transition at a critical separation. There are no orbits that are sometimes librating and sometimes
circulating.

Around an eccentric orbit binary, the secular tilt oscillations are driven 
by binary-planet interactions, as well as planet-planet interactions. An eccentric
orbit binary results in an additional planet tilt frequency. 
Due to this additional frequency, the planet tilt oscillations are generally not periodic  (see Figures~\ref{aA1} to \ref{aA3}),
since the sum of periodic functions is not generally periodic.
The transition from mutual planet libration to circulation is not sharp and there
is a range of separations for which the planet orbits are neither purely librating nor purely circulating.
Instead, such orbits are sometimes librating and sometimes circulating (see Figure~\ref{aA3}). 
The range of planet separations over which both librating and circulating orbits occur increases with binary eccentricity (see Figure~\ref{cri}).
In addition, at certain separations, there are resonances for which the tilt oscillations are less complicated and periodic (see Figure~\ref{aT1}). 
Such resonances occur when the ratio of the contributing tilt frequencies is an integer.

In the case that the planets are nearly coplanar with respect to the binary  plane, the only stationary (nonoscillatory) tilt
configuration occurs when the planet orbital planes are aligned with the binary orbital plane.
For planets that are highly misaligned with respect to an eccentric orbit binary, there are
stationary tilt configurations in a frame that precesses with the binary that are generalisations 
of polar configurations for the single planet case (Figure~\ref{iphase}). The companion planet can lead to a large change
in the tilts required for the stationary configuration to tilts that are substantially less than $90^\circ$ (see Figure~\ref{istat}). 
In the limit of small planet masses, the
level of misalignment between planet orbits in a stationary polar
configuration increases with planet mass, even though their
level of mutual interaction increases. The reason is that the
planet interactions with the binary cause the planet stationary
tilt angles to change differently.
Tilt oscillations occur for initial departures from the stationary
configuration. Since the stationary tilts for the two polar planets generally differ, planets that begin in a mutually coplanar polar configuration generally undergo  
tilt oscillations (see Figure \ref{H1H2}).

The analytic model provides physical insight and confirmation of the numerical results.
The analytic model for the nearly coplanar case generally agrees well with the numerical simulations for small initial planet inclination $i_0=1^\circ$  (e.g.,  Figures~\ref{aA0} to \ref{aC2}).
The agreement is somewhat less good for larger inclinations of $i_0=10^\circ$, as is expected since the analytic model
assumes the tilts are small.  The analytic model for the polar stationary angles agrees well with simulations (Figure~\ref{istat}). But the agreement for the
tilt oscillations breaks down after a significant tilt change occurs (see Figure \ref{H1H2}). This breakdown is likely a consequence of the high level
of misalignment between the outer planet and the polar state and the level of misalignment between the planets.

Orbital stability is an important issue for multi-planet circumbinary systems. The stability of a single, coplanar, close-in circumbinary planet for different binary parameters has been investigated by previous work \citep{Holman1999, Popova2016}. In \citet{Chen20201}, we extended their work for different values of the planet inclination and found that the the planet's orbital stability is affected by the planet orbital inclination, as well as planet semi-major axis, binary mass ratio, binary eccentricity, and planet mass. However, planet-planet interactions can additionally destabilise the planet's orbit and may easily eject a planet from multi-planet systems both around single or binary stars \citep{Davies2014}. Resonances between two CBPs result in orbital migration and instability \citep{Sutherland2019}. Simulations of pre-main sequence binary systems that undergo tidal and magnetic breaking effects suggest that multi-planet circumbinary systems often eject a close-in circumbinary planet \citep{Fleming2018}. In addition, a circumbinary planet may be ejected from the system during its inward migration due to interaction with a circumbinary disc \citep{Kley2014,Kley2015}.


We considered a single value for the planet mass in this paper.
The mass of the planet affects the range of separations over which
the effects analysed in this paper operate. For example, for lower
mass planets the critical separation for the transition from libration
to circulation is expected to decrease due to the reduction in the
strength of planet-planet interactions. Furthermore, the stationary tilt
angles for planets should get closer to $90^\circ$ for reduced planet masses.
For higher planet angular momentum values, the binary orbit will undergo stronger
orbital element variations \citep{Chen20192}.
However, we expect the planet orbital transitions to often operate in a qualitatively similar manner to what we find in this paper.

There are other effects we have not considered.
If the outer planet is massive enough, it could trigger evection resonance when the apsidal precession frequency of the inner planet matches the orbital frequency of the outer planet. It results in inward migration and eccentricity excitation of the inner planet and it may cause a collision with the binary or ejection from the system \citep{Xu2016}. Also, if two planets are located far from the binary and the relative inclinations of planets are high, they may undergo Kozai-Lidov oscillations because the binary can be considered to an point mass \citep{Kozai1962, Lidov1962}. During Kozai-Lidov oscillations, the eccentricity of the inner planet could be excited to very high values, resulting in a close encounter of the inner planet with the central binary. Consequently, the inner planet may be flung out to a larger radius or collide with the binary.

Several binary systems have been found with highly misaligned circumbinary discs. KH 15D is a young binary system which consists of a misaligned circumbinary disc (3 $\sim$ 15$^\circ$) that was found using spectroscopic and photometric data \citep{Chiang2004,Poon2020}. GW Ori is a triple star system that includes a misaligned circumtriple disc. With ALMA observations, three dust rings which have different mutual inclinations (> 10$^\circ$) to the stars have been identified \citep{Bi2020, Kraus2020}. An initially misaligned disc that is not at a stationary inclination around an eccentric orbit binary undergoes tilt oscillations, even if it is evolving towards coplanarity \citep{Smallwood2019}, and may evolve towards a more misaligned (polar) configuration with respect to the binary, such as in HD98800 \citep{Kennedy2019}. Planets formed in such discs will be on misaligned orbits with respect to the binary orbital plane. Because two circumbinary planets in the same system have different stationary inclinations, they may evolve to different tilts even if they were coplanar to each other initially. Therefore, we expect multi-planet circumbinary systems discovered in the future will display diverse orbital configurations.


As discussed in the Introduction, the Kepler and TESS missions have detected circumbinary planets and more will likely be found.
It may be possible to find highly inclined circumbinary planets
with TESS and other missions using eclipse timing variations of the binary (ETVs) \citep{Zhang2019}.
Moreover, the PLATO mission will monitor nearly 1,000,000 stars to search transits and new circumbinary planets will likely be found \citep{Rauer2014}.
The mutual inclinations of high mass circumbinary planets and binaries may be determined by the Gaia mission \citep{Sahlmann2015}. Such results will be a key component for understanding planet-planet and planet-binary oscillations.

\section*{Acknowledgements}
Computer support was provided by UNLV's National Supercomputing Center. C.C. acknowledges support from a UNLV graduate assistantship. We acknowledge support from NASA through grants 80NSSC21K0395 and 80NSSC19K0443. Simulations in this paper made use of the REBOUND code which can be downloaded freely athttp://github.com/hannorein/rebound.

\section*{Data availability}
The data underlying this article will be shared on reasonable request to the corresponding author.



\bibliographystyle{mnras}
\bibliography{main} 

\bsp	
\label{lastpage}
\end{document}